\documentclass[
reprint,
nofootinbib,
amsmath,amssymb, aps
prd,
]{revtex4-2}
\pdfoutput=1

\usepackage{graphicx}
\usepackage{color}
\usepackage{amsmath}

\newcommand{\beq}{\begin{equation}}
\newcommand{\eeq}{\end{equation}}
\newcommand{\bea}{\begin{eqnarray}}

\newcommand{\eea}{\end{eqnarray}}
\newcommand{\vc}[1]{{\textbf{#1}}}
\newcommand{\bmatr}{\begin{pmatrix}}
	\newcommand{\ematr}{\end{pmatrix}}

\numberwithin{figure}{section}
\begin{document}

\title{$\Delta\mathcal{N}$ and the stochastic conveyor belt of Ultra Slow-Roll}

\author{Tomislav Prokopec}
\affiliation{Institute for Theoretical Physics, Spinoza Institute and the
	Center for Extreme Matter and Emergent Phenomena ,
	Utrecht University, Buys Ballot Building,
	Princetonplein 5, 3584 CC Utrecht, the Netherlands}
\author{Gerasimos Rigopoulos}
\affiliation{School of Mathematics, Statistics and Physics, 
	Herschel Building, Newcastle University, 
	Newcastle upon Tyne, NE1 7RU, UK}

\begin{abstract}
We analyse field fluctuations during an Ultra Slow-Roll phase in the stochastic picture of inflation and the resulting non-Gaussian curvature perturbation, fully including the gravitational backreaction of the field's velocity. By working to leading order in a gradient expansion, we first demonstrate that consistency with the momentum constraint of General Relativity prevents the field velocity from having a stochastic source, reflecting the existence of a single scalar dynamical degree of freedom on long wavelengths.  We then focus on a completely level potential surface, $V=V_0$, extending from a specified exit point $\phi_{\rm e}$, where slow roll resumes or inflation ends, to $\phi\rightarrow +\infty$. We compute the probability distribution  in the number of e-folds $\mathcal{N}$ required to reach $\phi_{\rm e}$ which allows for the computation of the curvature perturbation. We find that, if the field's initial velocity is high enough, all points eventually exit through $\phi_{\rm e}$ and a finite curvature perturbation is generated. On the contrary, if the initial velocity is low, some points enter an eternally inflating regime despite the existence of $\phi_{\rm e}$. In that case the probability distribution for $\mathcal{N}$, although normalizable, does not possess finite moments, leading to a divergent curvature perturbation.
\end{abstract}
\maketitle


\section{Introduction}
\label{Introduction}
	
The $\Delta \mathcal{N}$ formalism is a very convenient way to compute the curvature perturbation generated during inflation. Its basic tenet is that quantum fluctuations stretched to superhorizon scales introduce randomness in the total number of e-folds at different points in the universe, with this number of e-folds counted from a given initial spatially flat time-slice to a given final uniform $\phi$ time-slice. This final time-slice is determined by a prescribed condition on the scalar field, for example that slow-roll, and presumably inflation, ends. This difference in the number of e-folds between different spatial points directly gives the scalar curvature perturbation. The idea that a time delay encodes the curvature perturbation induced by the fluctuations of the inflaton goes back to the early days of inflationary cosmology and was already used in some of the pioneering papers on inflationary pertubations~\cite{Hawking:1982cz,Starobinsky:1982ee}. The relation of classicalized, super-Hubble modes to a time delay in the dynamics was clearly explained in ~\cite{Guth:1985ya}. The concept was further formalized and connected to the conserved gauge invariant curvature perturbation in~\cite{Salopek:1990jq} and later in~\cite{Sasaki:1995aw, Sasaki:1998ug}. More recently, it was re-introduced and elaborated in~\cite{Lyth:2004gb, Lyth:2005fi}, see also e.g.~\cite{Sugiyama:2012tj,Garriga:2015tea}. 

In the usual slow roll scenario, the number of e-folds between the prescribed time slices is dominated by the classical/deterministic result and random perturbations are introduced only as a fluctuation of the initial condition in $\phi$, generated when a given mode exits the Hubble radius. In this regime, the evolution of the probability distribution is dominated by the drift term of the Fokker-Planck equation. However, when the potential is very flat, as in the case of ultra slow roll \cite{Tsamis:2003px, Kinney:2005vj, Martin:2012pe,Namjoo:2012aa, Dimopoulos:2017ged}, and the drift term is small, the slow-roll formula for the curvature perturbation cannot be used any more and the e-fold number becomes an essentially stochastic quantity. When asking for the time it takes for the field to reach the prescribed value $\phi_{\rm e}$ we are thus facing a \emph{first-passage time} problem in the stochastically evolving system: Given an initial condition, how many e-folds $\mathcal{N}$ are needed for $\phi_{\rm e}$ to be reached? The total number of e-folds becomes a stochastic quantity described by a probability distribution $\varrho(\mathcal{N})$, such that $\varrho(\mathcal{N})d\mathcal{N}$ is the probability that the field will reach $\phi_{\rm e}$ for the first time within the interval $[\mathcal{N},\mathcal{N}+d\mathcal{N})$ of e-folds~\cite{Starobinsky:1986fx,Salopek:1990re, Enqvist:2008kt,Fujita:2013cna,Fujita:2014tja,Vennin:2015hra}. As alluded to above, one can define the first passage with respect to any desirable condition labelled by $\phi_{\rm e}$ and defined by a constant field hypersurface where $\phi=\phi_{\rm e}$, such as inflation ending or the commencement of another distinct phase.   
	
The curvature perturbation generated during a phase of Ultra Slow Roll (USR) has attracted considerable attention recently~\cite{Germani:2017bcs,Pattison:2017mbe, Ezquiaga:2018gbw, Biagetti:2018pjj, Firouzjahi:2018vet, Cruces:2018cvq,Passaglia:2018ixg,Pattison:2019hef} due to the possibility that it leads to an enhanced curvature perturbation and a related enhanced primordial black hole production - see \cite{Garcia-Bellido:2018leu} for a recent review on cosmological implications of primordial black holes. In this work we revisit the problem,  taking the scalar sector of gravity fully into account. We place the computation within the framework of a long wavelength (leading gradient) approximation to the equations of General Relativity for an inhomogeneous universe: we retain full non-linearities but drop terms that are second order in spatial gradients and properly take into account the field's velocity and the corresponding gravitational backreaction. Quantum fluctuations are then consistently included as a random forcing of the dynamical equation of the scalar field, a well established approximation for IR quantum fields in inflationary spacetimes \cite{Starobinsky:1986fx, Tsamis:2005hd,Finelli:2008zg,Finelli:2010sh,Garbrecht:2013coa,Garbrecht:2014dca,Moss:2016uix}. We find that imposing the $0i$ Einstein equation, the GR momentum constraint, leads to the field $\phi$ being the only dynamical stochastic variable. The field's velocity is constrained and does not obey an independent equation involving different stochastic kicks at different spatial points, unlike what a na\^ive ``separate universe'' argument would imply. This is a non-linear generalization of the linear perturbation theory result for $k/aH \rightarrow 0$ \cite{Salopek:1990jq, Rigopoulos:2005us}.     

We apply the formalism to a simple problem: the curvature fluctuation generated in an extreme version of USR where the field is injected at some point $\phi_{\rm in}$ with velocity $\Pi_{\rm in}$ on a totally flat potential $V=V_0$. We find two separate regimes, depending on the distance from $\phi_{\rm in}$ to the exit point $\phi_{\rm e}$ and the initial velocity. If this distance is  
larger than the length of the classical trajectory, corresponding to a low injection velocity $\Pi_{\rm in}$,
the field experiences what we call the \emph{stochastic conveyor belt model} for USR: in some parts of the universe the initial velocity is forgotten and the field  explores the infinite semi-line $\phi \rightarrow \infty$, never fully reaching $\phi_{\rm e}$. The resulting probability distribution for the total number of e-folds $\mathcal{N}$ is normalizable but does not have finite moments, leading to the infinite inflation observed in \cite{Assadullahi:2016gkk, Vennin:2016wnk}. However, if the distance is smaller than the length of the classical trajectory, corresponding to a high injection velocity $\Pi_{\rm in}$,
graceful exit does occur and inflation eventually terminates in all points of the Universe. The curvature perturbation is then finite and is described by a highly non-Gaussian probability 
distribution that we compute. 

Obviously, the infinite inflation regime will not be reached in realistic single field models where USR takes place only on a finite portion of the potential. This paper then serves an expository function for the developed techniques, involving mainly the use of the Hamilton-Jacobi equation for inflationary evolution, the consistent inclusion of stochastic fluctuations, and the description of the stochastic conveyor belt mechanism. These techniques will be used to analyse more realistic USR potentials in a forthcoming publication \cite{Rigopoulos:2021nhv}.


\section{Long wavelength scalar perturbations}
\label{Hamilton-Jacobi approach to long wavelength scalar perturbations}

We start by recalling the long wavelength approach of Ref.~\cite{Salopek:1990jq}, 
see also~\cite{Rigopoulos:2003ak}, which will take us to the starting point of our stochastic analysis. 
Considering the metric in its ADM parametrization, 
\beq
g_{00} = - N^2 + \gamma^{ij}N_iN_j\,,\quad g_{0i} = N_i\,,\quad g_{ij}=\gamma_{ij}\,,
\eeq 
where $N$ and $N_i$ are the lapse function and shift vector  respectively,
the Einstein equations for gravity plus a single scalar field $\phi$ give the GR energy and momentum constraints ($00$ and $0i$ Einstein equations)
\begin{eqnarray}
\bar{K}_{ij}\bar{K}^{ij}-\frac{2}{3}K^2-^{(3)}\!\!\!R + 16\pi G \varepsilon = 0 \,, \label{energy constraint:0}\\
\bar{K}^j_{i|j} -\frac{2}{3}K_{|i}+8\pi G\,\Pi\,\phi_{|i}=0\,,  
\label{momentum constraint:0}
\end{eqnarray}  
the dynamical equations for the extrinsic curvature tensor of the 3-slices $K_{ij}=\bar{K}_{ij}+\frac{1}{3}K\gamma_{ij}$
\begin{eqnarray}\label{dKt}
\frac{\partial K}{\partial t} &-& N^i K_{|i} = -N^{|i}{}_{|i}\nonumber\\ &+& N \left[\frac{3}{4}\bar{K}_{ij}\bar{K}^{ij} + \frac{1}{2}K^2 + \frac{1}{4}{}^{(3)}\!R+4\pi G S\right]\,,
\end{eqnarray}
\begin{eqnarray}\label{dkbart}
\frac{\partial \bar{K}^i{}_k}{\partial t} &+& N^i{}_{|l} \bar{K}^l_k-N^l{}_{|k}\bar{K}^i_l - N^l\bar{K}^i_{k|l} = -N^{|i}{}_{|k}\nonumber\\
 &+&\frac{1}{3} N^{|l}{}_{|l}\delta^i_k  + N \left[K\bar{K}^i_k  + ^{(3)}\!\!\!\bar{R}^i_k-8\pi G \bar{S}^i_k\right]\,,
\end{eqnarray}
(stemming from the $ij$ Einstein equation) and the equation of motion for the scalar field 
\begin{equation}
\frac{1}{N}\left(\frac{\partial \Pi}{\partial t} - N^i\Pi_{|i}\right) - K\Pi - \frac{1}{N}N_{|i}\phi^{|i} - \phi_{|i}\phi^{|i} + \frac{d V}{d\phi}=0  \,.
\label{field equation}
\end{equation}
In the above, the field momentum $\Pi$ is defined as
\begin{equation}
\Pi=\frac{1}{N}\left(\frac{\partial \phi}{\partial t}-N^i\phi_{|i}\right)
\,,
\label{scalar field momentum}
\end{equation} 
the extrinsic curvature 3-tensor is 
\begin{equation}
K_{ij} = -\frac{1}{2N}\left(\frac{\partial \gamma_{ij}}{\partial t}-N_{i|j}-N_{j|i}\right)\,,
\end{equation}
and the scalar's energy density and stress tensor on the 3-slices read
\begin{eqnarray}
\varepsilon = \frac{1}{2}\left(\Pi^2+\phi_{|i}\phi^{|i}\right) + V(\phi)\,,
\end{eqnarray}
and
\begin{equation}
S_{ij} = \phi_{|i}\phi_{|j}+\gamma_{ij}\left(\frac{1}{2}\Pi^2-\frac{1}{2}\phi_{|i}\phi^{|i}-V(\phi)\right)\,. 
\end{equation} 
A vertical bar denotes a covariant derivative {\it w.r.t.}~the 3-metric $\gamma_{ij}$ which is also used to raise or lower spatial indices. 

The approximation we use to study the non-linear long wavelength configurations relevant for inflation is to only keep terms containing the leading order in spatial derivatives. This is underpinned by the expectation that on scales $aL>H^{-1}$ the dynamics is dominated by time derivatives such that for any quantity $Q$ the inequality 
$\|e^{-\alpha}\nabla Q\| <|\partial_t Q|$ will be true, 
where $\dot\alpha$ denotes the local expansion rate 
(see Eqs.~(\ref{3-metric}) and (\ref{local hubble}) below),
a statement that in inflation is expected to eventually hold for all scales of interest.  
Furthermore, to simplify the equations we choose to consider coordinate systems constructed such that
$N_i=0$. This gauge choice fixes three gauge degrees of freedom, 
leaving one gauge function unfixed; its elimination can be achieved {\it e.g.} 
by further choosing a specific form for the lapse function $N$.
 Under these assumptions, we get from~(\ref{dkbart}) that the traceless part of the extrinsic curvature evolves according to 
\begin{equation}
\frac{\partial \bar{K}^i{}_k}{\partial t} = N K\bar{K}^i_k \,.
\end{equation}   
The 3-metric can be further decomposed as   
\begin{equation}
\gamma_{ij} = e^{2\alpha}h_{ij}
\label{hij def}
\end{equation}
where ${\rm Det}\,h_{ij}=1$ and therefore $h^{ij}\frac{\partial}{\partial t} h_{ij}=0$. We then have 
\begin{equation}
K=-3\dot{\alpha} \equiv -3\frac{1}{N}\frac{\partial\alpha}{\partial t}
\,,
\end{equation}  
from where we directly obtain 
\begin{equation}
\bar{K}^i{}_j = C^i{}_j(\mathbf{x})e^{-3\alpha}
        \,,\qquad  {\rm Tr}[C^i{}_j(\mathbf{x})]=0
\,.
\label{anisotropic expansion rate: solution} 
\end{equation}
Since during inflation $\alpha(t,\mathbf{x})$ represents the local generalization of the number 
of e-folds, it grows approximately linearly in time, and we can take it as a proxy for {\it time}
in inflation. 
Therefore, Eq.~(\ref{anisotropic expansion rate: solution}) tells us that 
 the anisotropic expansion rate $\bar{K}^i{}_j$ -- which 
 is the non-linear generalization of the canonical momentum associated with gravitational waves
 -- declines extremely rapidly (exponentially fast) 
during inflation. We are thus dynamically led to $\bar{K}^i{}_j =0$ and the most general 3-metric on long wavelengths can be written as 
\begin{equation}\label{3-metric}
\gamma_{ij}(t,\mathbf{x}) = e^{2\alpha(t,\mathbf{x})}h_{ij}(\mathbf{x})
\end{equation}
with the long wavelength spacetime metric taking the form  
\begin{equation}\label{ADM scalar}
ds^2=-N^2(t,\mathbf{x})dt^2+e^{2\alpha(t,\mathbf{x})}h_{ij}(\mathbf{x})dx^idx^j
\,.
\end{equation}
and the 3-tensor $h_{ij}$ is not dynamical in this approximation, at least classically. 
Furthermore, we restrict the lapse function $N(t,\mathbf{x})$ to vary slowly enough in space
such that its spatial gradients can be neglected.
Later on we will consider scalar quantum fluctuations and the accompanying tensor fluctuations would provide $h_{ij}$ with a stochastic source from subhorizon tensor modes entering the long wavelength sector and with an amplitude set by the uncertainty principle. In this work we focus on the dynamics of the scalar sector of gravity, leaving that of the stochastic evolution of the tensor sector for future study. 

Defining the local expansion rate as 
\beq\label{local hubble}
H(t,\vc{x})\equiv\frac{1}{N}\frac{\partial \alpha}{\partial t}
\,,
\eeq     
we are led to a set of long wavelength equations for the spatially dependent field $\phi(t,\mathbf{x})$ and expansion rate $H(t,\mathbf{x})$ comprising of 
the two GR constraints: the energy~(\ref{energy constraint:0}) 
and momentum~(\ref{momentum constraint:0}) constraint (from now on we set $8\pi G =1$~\footnote{The Newton
constant $8\pi G=1/M_{\rm P}^2$ can always be recovered in the equations by noting 
the canonical dimension of various quantities, 
$[H]=1, [\phi]=1, [8\pi G]=-2, [\Pi]=2, [V(\phi)]=4, [h_{ij}]=0$.})
\beq
H^2=\frac{1}{3}\left(\frac{1}{2}\Pi^2+V(\phi)\right)
\label{Energy-constraint}
\eeq
and
\beq
\partial_iH=-\frac{1}{2}\Pi\partial_i \phi \,,
\label{H-gradient-constraint}
\eeq
the evolution of the expansion rate~(\ref{dKt}),
\begin{equation}\label{H-evolution}
\frac{1}{N}\frac{\partial H}{\partial t} = -\frac{1}{2}\Pi^2\,,
\end{equation}
as well as the dynamical equations
 for the scalar field~(\ref{field equation}--\ref{scalar field momentum}),
\beq\label{Pi-defn}
\Pi=\frac{1}{N}\frac{\partial \phi}{\partial t}
\eeq
\beq\label{Pi-dot}
\frac{1}{N}\frac{\partial \Pi}{\partial t}+3H\Pi+\frac{dV}{d\phi}=0
\,.
\eeq

Equations (\ref{Energy-constraint}), (\ref{H-evolution}), (\ref{Pi-defn}) and (\ref{Pi-dot}) are formally the same as those of homogeneous cosmology but are valid at each spatial point with a priori different values of the initial conditions for $\phi$ and $\Pi$. This is what is sometimes referred to as the ``{\it separate universe evolution}''. However, not all spatially inhomogeneous initial conditions are allowed on long wavelengths as they must also satisfy the momentum constraint (\ref{H-gradient-constraint}). As we will see, this restricts the possibility of assigning $\Pi$ independently of the initial value of $\phi$. Indeed, for the constraint (\ref{H-gradient-constraint}) to be respected the local expansion rate $H$ must depend on the spatial position only through its dependence on the spatially varying $\phi$
\begin{equation}\label{H-of-phi}
H(t,\vc{x})=H(\phi(t,\vc{x}),t)
\end{equation}
and the field momentum must be given by  
\beq\label{Pi-as-gradient}
\Pi=-2\frac{\partial H}{\partial \phi}
\,.
\eeq
Taking the time derivative of (\ref{H-of-phi}) and comparing with (\ref{H-evolution}) we immediately find that   
\begin{eqnarray}
\left(\frac{\partial H}{\partial t}\right)_\phi =0\,,
\end{eqnarray}
showing that the total spatio-temporal dependence of the expansion rate is solely determined through its dependence on $\phi(t,\vc{x})$
\beq
H=H(\phi(t,\vc{x}))\,,
\eeq
and the field momentum is therefore given by 
\begin{equation} \label{Pi-as-gradient-2}
\Pi=-2\frac{d H}{d \phi}\,.
\end{equation} 
Hence $\Pi$ is evidently also a function of $\phi$ alone, $\Pi=\Pi(\phi(t,\mathbf{x}))$, with no explicit temporal or spatial dependence. 

When (\ref{Pi-as-gradient-2}) is inserted into the local energy constraint (\ref{H-gradient-constraint}) it gives the 
{\it Hamilton-Jacobi equation} for the function $H(\phi)$ 
\beq\label{Hamilton Jacobi}
\left(\frac{dH}{d\phi}\right)^2=\frac{3}{2}H^2-\frac{1}{2}V(\phi)
\,.
\eeq 
Since (\ref{Hamilton Jacobi}) is a first order differential equation it admits a family of solutions of the form $H(\phi, \mathcal{C})$, with different solutions parametrized by an arbitrary constant $\mathcal{C}$. A solution $H=H(\phi,\mathcal{C})$ to~(\ref{Hamilton Jacobi}) along with 
\beq\label{phi-dynamics}
\frac{d\phi}{dt}= -2N\frac{d H}{d \phi}
\eeq 
provide a \emph{complete description} and the inhomogeneous long wavelegth scalar field configuration $\phi$.\footnote{Equation (\ref{Hamilton Jacobi}), often referred to as the Hamilton-Jacobi equation in the literature, has been used in \cite{Liddle:1994dx, Binetruy:2014zya} to study an alternative parameterization of possible homogeneous inflationary cosmologies, including USR \cite{Kinney:2005vj, Cicciarella:2017nls}. We stress that in the present, long wavelength context it describes an \emph{inhomogeneous} universe as well, as originally demonstrated in \cite{Salopek:1990jq}.} The corresponding long wavelength metric (\ref{ADM scalar}) can then determined via
\begin{equation}\label{alpha-dynamics}
\frac{1}{N}\frac{\partial \alpha}{\partial t}= H(\phi)\,.
\end{equation}

It should be pointed out that by taking a $\phi$ derivative of~(\ref{Energy-constraint}) one can verify that the long wavelength field indeed obeys
\begin{equation}\label{2nd order phi}
\frac{1}{N}\frac{\partial}{\partial t}\left(\frac{1}{N}\frac{\partial \phi}{\partial t}\right)+3H\frac{1}{N}\frac{\partial\phi}{\partial t}+\frac{dV}{d\phi}=0
\end{equation}
as expected. Although equations (\ref{Hamilton Jacobi}) - (\ref{alpha-dynamics}) are identical to those of homogeneous cosmology, including (\ref{2nd order phi}) which is implied by them, the momentum constraint (\ref{H-gradient-constraint}) imposes that there is no freedom to choose initial values for $\Pi$ independently at each spatial point. Let us see why: If (\ref{Hamilton Jacobi}) is considered without reference to (\ref{H-gradient-constraint}), a na\^ive separate universe picture would imply that $\mathcal{C}=\mathcal{C}(\vec x)$, {\it i.e.} for every point $\vec x$ on the initial hypersurface 
there would be a separate integration constant $\mathcal{C}(\vec x)$. This simply  encodes the freedom to choose the initial field momentum at that point independently of $\phi$. However, a spatially inhomogeneous $\mathcal{C}(\vec x)$ further leads to 
\begin{equation}
\nabla H =(\partial_{\mathcal{C}} H) \nabla \mathcal{C} +(\partial_\phi H)\nabla\phi 
\neq-\frac{1}{2}\Pi\nabla \phi
\,.\quad
\label{momentum constraint 2}
\end{equation}
Therefore, as long as $\partial_{\cal C} H \neq 0$ it follows that $\nabla \mathcal{C}=0$, otherwise the momentum constraint (\ref{H-gradient-constraint}) would be violated. This restricts $\mathcal{C}$ to be a global constant, meaning that the momentum $\Pi$ cannot be chosen arbitrarily at each point but, instead, all spatial points must be placed along one and the same integral curve of (\ref{Hamilton Jacobi}). 

The restriction $\nabla \mathcal{C}=0$ is a direct consequence of the momentum constraint. However, it would not apply when integral curves of $H(\phi,\mathcal{C})$ exist for which $\partial_{\mathcal{C}} H = 0$. In this case setting $\nabla\mathcal{C} \neq 0$ would not violate the momentum constraint since a spatially varying value of $\mathcal{C}$ does not change the value of $H$ at different spatial points. Such integral curves are attractors of the long wavelength system as one can see by taking a derivative of (\ref{Hamilton Jacobi}) w.r.t.~$\mathcal{C}$ which leads to \cite{Salopek:1990jq, Rigopoulos:2003ak} 
\begin{equation}\label{eq:decaying}
\frac{\partial H}{\partial \mathcal{C}} \propto a^{-3}\,.
\end{equation}     
This is equivalent to the decay of the second solution to (\ref{2nd order phi}) and, as was already pointed out in \cite{Sugiyama:2012tj}, if this decaying mode is neglected the momentum constraint places no further restriction on the long wavelength configuration; this is the case for example in Slow Roll inflation. In the case of motion on a constant potential examined in this work, the decaying mode (\ref{eq:decaying}) expresses the approach to a static field and de Sitter spacetime - see (\ref{general solution of HJ: USR2}) where the constant $\phi_0$ in that formula plays the role of the general constant $\mathcal{C}$ discussed here. Including this decaying mode is therefore crucial in studying the slide and slowing down of the field on the constant potential. More generally, the Hamilton-Jacobi treatment adopted here allows the $\Delta \mathcal{N}$ formalism to be made fully and a priori consistent with the momentum constraint and include the decaying mode which is essential in studying Ultra Slow Roll. 

We can conclude that on sufficiently large scales, at which subleading spatial gradient terms can be neglected, the dynamics of the inhomogeneous configuration can be described solely in terms of $\phi$ while the momentum, when it contributes to the expansion rate, cannot be arbitrarily chosen at different spatial points. The would-be inhomogeneous degree of freedom is killed by the momentum constraint~(\ref{H-gradient-constraint}) on long wavelengths which patches different spatial points together. Physically, that means that on large scales there is a decaying inhomogeneous mode which is necessarily suppressed by spatial derivatives, and hence negligible in the leading gradient expansion.
We stress that this conclusion goes beyond slow roll and is completely general, relying only on the long wavelength approximation. Note that slow-roll trivially satisfies the momentum constraint and is an attractor in which any dependence of $H$ on $\mathcal{C}$ is altogether suppressed exponentially.
The absence of a second dynamical inhomogeneous mode is unique to single scalar inflationary models. For a recent study on how the inflaton canonical momentum can be excited during inflation through its coupling to a light spectator scalar field, see Ref.~\cite{Friedrich:2019hev}.

Before closing this section we note that equations (\ref{Hamilton Jacobi}), (\ref{phi-dynamics}) and (\ref{alpha-dynamics}) and the corresponding metric (\ref{ADM scalar}) are valid for \emph{any} choice of time hyper-surfaces (any choice of $N$) as long as the corresponding spatial coordinate worldlines are constructed orthogonal to the time slices, keeping $N_i=0$, and if all terms that are second order in spatial gradients are dropped, see \cite{Salopek:1990jq}. Fer the reader's convenience, we recall the demonstration of this fact in Appendix A.


\section{Hamilton-Jacobi solution in Ultra Slow-Roll}
\label{Ultra Slow Roll}

We now apply the above analysis to our extremal USR scenario where the field moves along a very flat and level part of the potential where $V(\phi)\simeq V_0$. Equation (\ref{Hamilton Jacobi}) then reads
\beq\label{USR Hamilton Jacobi}
\left(\frac{dH}{d\phi}\right)^2=\frac{3}{2}H^2-\frac{1}{2}V_0
\,.
\eeq
Taking a derivative with respect to $\phi$ gives, 
\begin{equation}
\frac{dH}{d\phi}\left(\frac{d^2H}{d\phi^2}-\frac32H\right)=0
\,,
\label{derivative of HJ equation:USR}
\end{equation}
and the general solution can be written as
\begin{equation}
 H(\phi) = H_0=\sqrt{\frac{V_0}{3}}
 \,,\quad {\rm \underline{or}}\quad
   H(\phi)=A{\rm e}^{-\sqrt{\frac{3}{2}}\phi}+B{\rm e}^{\sqrt{\frac{3}{2}}\phi}
\label{general solution of HJ: USR}
\end{equation}
where $AB = \frac{V_0}{12}$. The constraints on the integration constants $H_0$ and $AB$ are obtained by 
inserting  the general solutions of~(\ref{derivative of HJ equation:USR}) into 
the original Hamilton Jacobi equation~(\ref{USR Hamilton Jacobi}).
It is convenient to redefine $A$ and $B$ as
\begin{equation}
A=\frac{H_0}{2}{\cal C}=\frac{H_0}{2}{\rm e}^{\sqrt{\frac{3}{2}}\phi_0}
\,,\quad 
B=\frac{H_0}{2}{\cal C}^{-1}=\frac{H_0}{2}{\rm e}^{-\sqrt{\frac{3}{2}}\phi_0}
\,,
\label{A and B}
\end{equation}
where ${\cal C}$ and $\phi_0$ are global constants. With these definitions in mind,
the second solution in~(\ref{general solution of HJ: USR}) becomes, 
\begin{eqnarray}
 H(\phi) &=& \frac{H_0}{2}\left({\cal C}{\rm e}^{-\sqrt{\frac{3}{2}}\phi}
                     +{\cal C}^{-1}{\rm e}^{\sqrt{\frac{3}{2}}\phi}\right)\nonumber\\
          &=& H_0\cosh\left(\sqrt{\frac{3}{2}}(\phi-\phi_0)\right)
   \,,
\label{general solution of HJ: USR2}
\end{eqnarray}
where we replaced ${\cal C}$ by $\phi_0=\sqrt{\frac{2}{3}}\ln({\cal C})$.
The corresponding momentum (velocity) is then from~(\ref{Pi-as-gradient}),
\beq
\Pi=0
\,,\quad {\rm \underline{or}}\quad
\Pi =-\sqrt{6}H_0\sinh\left(\sqrt{\frac{3}{2}}(\phi-\phi_0)\right)
\,.
\label{canonical momentum: USR}
\eeq
The number of e-foldings $\alpha$, from~(\ref{local hubble}) and using the solution
for $H(\phi)$~(\ref{general solution of HJ: USR2}) becomes
\begin{equation}
\alpha=\int HNdt = -\frac16\ln\left[\sinh^2\left(\sqrt{\frac{3}{2}}(\phi-\phi_0)\right)\right]
\,.
\label{number of e-foldings}
\end{equation}

\begin{figure}[t]
	\begin{center}
		\includegraphics[scale=0.45]{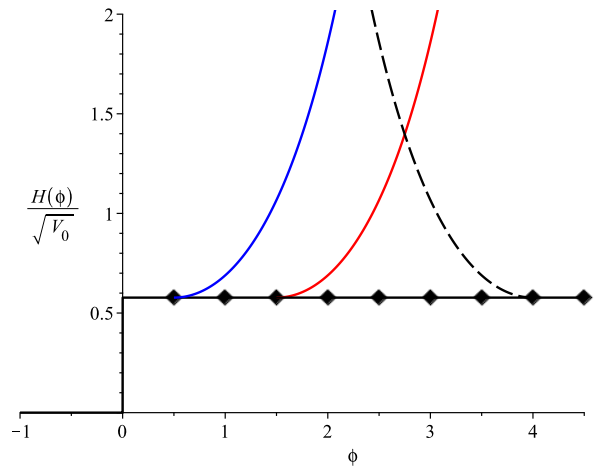}
		\caption{The conveyor belt of stochastic Ultra Slow-Roll: Solutions of the HJ equation of motion on a flat potential $V(\phi)=V_0$ are plotted. Inflation ends at $\phi=0$ in this figure. The blue and red curves are HJ solutions for fields moving to the left (negative initial velocity) while the dashed curve shows a trajectory that started off with positive initial velocity. These evolving trajectories reach the $H^2_0=V_0/3$ surface only  asymptotically after an infinite number of e-folds. Alternatively, the field can remain stationary at some value of $\phi$ on one of the $H^2_0=V_0/3$ points - a sample of them is denoted in the figure by diamonds. Classically, the field does not transition from an evolving state to a stationary state and therefore the classical phase space is partitioned into these two types of trajectories. Stochastic fluctuations disintegrate this partition: the field can start along one of the HJ trajectories but it is now possible to cross the asymptotic end point due to a stochastic jump. This end point acts as a bifurcation point for the quantum phase-space, beyond which the field simply diffuses along the $H^2=V_0/3$ surface. The GR momentum constraint is still respected since $\Pi=0$ there. The system thus resembles a jiggly conveyor belt where an initial HJ trajectory feeds into a de Sitter stage with successive unimpeded diffusion.}
	\end{center}
\label{figure1}
\end{figure}

The above formulae completely describe the classical
field evolution along a flat potential $V=V_0$, including the gravitational backreaction of a non-zero field velocity. However, they require some clarification with regards to the exact dynamics they describe. We see that if the field starts off with a finite velocity, it evolves asymptotically towards $\phi_0$, which it reaches only after an infinite amount of e-folds. The precise value of $\phi_0$ depends on the initial velocity imparted on $\phi$ as well as its sign: If $\Pi(\phi_{\rm in})>0$  then $\phi_0>\phi(t)$ and $\phi$ moves asymptotically to the right towards $\phi_0$. If $\Pi(\phi_{\rm in})<0$  then $\phi_0<\phi(t)$ and $\phi$ moves asymptotically to the left towards $\phi_0$. Of course, if $\Pi(\phi_{\rm in})=0$ then the field remains static and these are degenerate ``trajectories'' where $\Pi=0$ and $H=H_0$ always. Note that they represent distinct solutions of (\ref{Hamilton Jacobi}) and the field does not transition from a $\Pi\neq 0$ to a $\Pi=0$ state during its classical evolution, nor does its velocity change sign. The constant $\phi_0$ 
is the asymptotic end-point of each classical trajectory and it 
parametrizes different integral curves of the HJ equation (\ref{Hamilton Jacobi}); it is identified with the constant $\mathcal{C}$ of the general discussion in section 2. Note that $\partial_{\phi_0}H|_{\phi=\phi_0}=0$ and therefore a static field can become inhomogeneous without violating the momentum constraint. Figure~3.1 summarises the different solutions. We therefore see that on long wavelengths and for a flat level potential: 
\begin{itemize}
	\item An \emph{evolving} ($\Pi\neq 0$) long wavelength field configuration always tends
	at asymptotically late times to 
	the \emph{same} field value at all spatial points, $\phi(t,\vc{x}) \rightarrow \phi_0$.  
	
	\item An arbitrary \emph{inhomogeneous} field configuration is only allowed for a \emph{static} field, since $\partial_{\phi_0}H|_{\phi=\phi_0}=0$ in this case. This is a non-linear version of the growing mode of linearized pertubations
\end{itemize}          
As emphasized above, these represent two different solutions which classically do not evolve into each other.
In the following section we discuss how this picture changes when quantum effects are included. 


\section{Stochastic evolution: the conveyor belt of Ultra Slow Roll}\label{Conveyor Belt}

Let us now incorporate a stochastic element in the evolution, modelling as usual quantum fluctuations stretched to long wavelengths. We do not need to specify the amplitude of the noise terms at this point so we keep them general for this discussion. Generically, any stochastic "add-on" to the dynamics, regardless of the microscopic origin of the extra noise terms, can be thought of as adding an extra stochastic ``kick'' to the classical drift determined by the dynamical equations. Suppose we introduce noise in both the field and its momentum: in a discretized form of the time evolution their values would be updated after a time step $\Delta t$ as      
\begin{eqnarray}
&&\Delta \phi= -2\frac{\partial H}{\partial \phi} N \Delta t+\xi_\phi N\Delta t\,,\nonumber\\
&&\langle \xi_\phi(t) \xi_\phi(t') \rangle = \mathcal{A}\delta(t-t')\,,
\end{eqnarray}
\begin{eqnarray}
&&\Delta{\Pi} =-3H\Pi N\Delta t + \xi_\Pi N\Delta t\,,\nonumber\\ &&\langle\xi_\Pi(t)\xi_\Pi(t')\rangle=\mathcal{B}\delta(t-t')\,.
\end{eqnarray}
To be consistent with the constraints, $H$ should be a solution to the Hamilton-Jacobi equation. For the case we are considering it is given by (\ref{general solution of HJ: USR2}) with either $\phi>\phi_0$ or $\phi<\phi_0$ depending on the sign of the velocity - see figure~3.1.

The energy and momentum constraints are not dynamical equations and therefore they are not to be accompanied by some form of an extra stochastic force. As is well known, the constraints are preserved by the classical dynamical evolution and any consistent stochastic extension of the dynamics should also preserve them while the stochastic kicks are incorporated - the stochastically updated field and momentum should respect them too. We can achieve this by elevating the constants appearing in the solution of the HJ equation, whose values parametrise different possible velocities for fixed field values, to stochastic variables. Let's now see how this can be done in our case.   

The constant $\phi_0$ in (\ref{canonical momentum: USR}), characterizing different HJ solutions, can be linked to the velocity since we can write 
\beq \label{phi_0}
\phi_0 = \phi -\sqrt{\frac{2}{3}} {\rm arc}\sinh\left(-\frac{\Pi}{\sqrt{2}V_0^{1/2}}\right)\,, 
\eeq
and the choice of $\Pi$ for fixed $\phi$ is reflected in $\phi_0$ which also defines the breadth of field values covered by motion on the flat potential given the initial $\Pi$ of the field. If many possible initial conditions for $\Pi$ are contemplated, then $\phi_0$ defines the different asymptotic resting points corresponding to different initial momenta $\Pi_{\rm in}$ for a given initial value of $\phi$.  
A stochastic change in $\Pi$ at fixed $\phi$ would correspond to the field changing the HJ curve along which it evolves and this can be accommodated by promoting $\phi_0$ to a stochastic variable which would change according to~\footnote{Note 
that we are using It\^o's calculus here~\cite{Damgaard:1987rr,Stochastic} and we therefore keep $\Delta\Pi^2$ terms to follow changes to order $\Delta t$. Other choices are possible along with corresponding calculi and the results are invariant since $\mathcal{A}$ and $\mathcal{B}$ are independent of $\phi$.
} 
\beq
\Delta\phi_0 = \Delta\phi + \frac{\Delta \Pi}{3H} + \frac{1}{18H^3}\frac{\partial  H}{\partial \phi}\Delta\Pi^2\,,
\eeq  
leading to
\beq\label{stochastic-phi0}
\Delta \phi_0 = \frac{N\mathcal{B}}{18H^3}\frac{\partial H}{\partial\phi}N \Delta t +\left(\frac{1}{3H}\xi_\Pi +\xi_\phi\right) N \Delta t
\,.
\eeq  
At face value this provides a stochastic equation for $\phi_0$ which would now take different values at different spatial points. However, as we stressed above, unless $\frac{\partial H}{\partial \phi_0}=-\frac{\partial H}{\partial\phi}=\frac{\Pi}{2}= 0$, $\phi_0$ should only take a global value if the momentum constraint is to be respected. This cannot be accommodated in (\ref{stochastic-phi0}) since, for any choice of $\xi_{\phi}$ and $\xi_\Pi$, $\phi_0$ necessarily develops inhomogeneities. Hence, although one could a priory allow for stochastic changes in the velocity through stochastically jumping between different HJ trajectories on top of the stochastic $\phi$ displacement, the momentum constraint prevents that if $\Pi\neq 0$.

We are thus led to conclude that as long as $\Pi\neq 0$ the whole long wavelegth universe can only be located on different points of a single HJ trajectory with the following stochastic equation: 
\begin{eqnarray}
\label{phi-stochastic}
&&\Delta \phi= -2\frac{\partial H(\phi,\phi_0)}{\partial \phi} N \Delta t+\xi_\phi N\Delta t\,,\nonumber\\
&&\langle \xi_\phi(t) \xi_\phi(t') \rangle = \mathcal{A}\delta(t-t')
\,,
\end{eqnarray}
with either $\phi >\phi_0$ or $\phi <\phi_0$, depending on the fixed sign of the momentum. Once stochastic evolution takes the field past $\phi_0$, memory of the initial velocity is lost and it simply diffuses by a free random walk on the flat potential surface $V_0$ obeying 
\beq\label{free random walk}
{\Delta \phi}=\xi_{\phi}(t,\vc{x}) N \Delta t\,,\quad
\langle \xi_\phi(t) \xi_\phi(t') \rangle = \mathcal{A}\delta(t-t')
\,.
\eeq  
In this regime the field does jump between different HJ trajectories,
{\it i.e.} different points on the $H=H_0$ surface. This is now allowed as these degenerate solutions are characterized by $\partial_{\phi_0}H|_{\phi=\phi_0}=\Pi(\phi_0)=0$. Hence the momentum constraint is not violated by the universe occupying different solutions at different spatial points and stochastically jumping between them, becoming a collection of classically static field values that carry no extra energy.  

Note that the quantum fluctuations have a remarkable effect. The classical phase-space, consisting of the set of trajectories that solve the HJ equation~(\ref{Hamilton Jacobi}) and which are shown in figure~3.1, 
is split into (a) regular (non-degenerate, HJ) trajectories, which are characterized by 
an initial momentum $\Pi_{\rm in}$, the corresponding
 field value, $\phi_{\rm in}=\phi(\Pi_{\rm in})$, and end at 
 $\phi_0=\phi_0(\Pi_{\rm in})$ at which $\Pi=0$; (b) degenerate
trajectories characterized by $\Pi_{\rm in}=0$ and an arbitrary field value $\phi_0$. 
The quantum phase-space is very different however. A typical quantum/stochastic trajectory consists of a classical HJ branch $\Pi_{\rm in}$, 
which ends at 
$\phi_0=\phi(\Pi_{\rm in})$, supplemented by the set of all degenerate trajectories,
($\Pi=0$, $\phi\in\mathtt{R}$). The point $\phi_0=\phi(\Pi_{\rm in})$
is a bifurcation point, at which the quantum trajectory splits into two branches: 
$\phi>\phi_0$ and $\phi<\phi_0$, see figure~3.1. 
This quantum phase space picture resembles a conveyor belt for the quantum field which starts at a point on one of the HJ branches, diffuses downwards towards 
the bifurcation point at $\phi_0$ and then continues diffusing along the set of points shown as the horizontal line $\Pi=0$, $H=H_0\equiv \sqrt{V_0/3}$ in figure~3.1. 

Despite field fluctuations being generated, the above picture does not carry with it a well-defined curvature perturbation. A corresponding curvature perturbation emerges only when an exit point $\phi_{\rm e}$ is specified, where either inflation ends or another inflationary era follows by exiting the region where $V(\phi) = V_0$. If $\phi_{\rm e}$ lies outside the HJ branch, we are faced with the stochastic conveyor belt and a double first passage-time problem: Firstly to transition from a $\Pi\neq 0$ solution onto the $H=H_0$ surface (non-stochastic evolution does not allow this) and secondly to exit the $H=H_0$ region by reaching $\phi_{\rm e}$. If $\phi_{\rm e}$ is reached within the HJ branch we have a standard first passage-time problem and the conveyor is not operational. We analyse these cases in sections \ref{section:No-graceful-exit} and \ref{graceful exit} respectively.


\section{The case $\phi_{\rm e}<\phi_0$: USR without graceful exit} 
\label{section:No-graceful-exit}

As we demonstrated above, the gravitationally consistent inclusion of velocity to the problem of diffusion on a flat potential leads naturally to a two stage process when $\phi_{\rm e}<\phi_0$: 1) All spatial points diffuse along a single branch of the HJ solution until $\phi_0$ is crossed and then 2) each point that has crossed $\phi_0$ diffuses independently along the level $V_0$ potential. We therefore need to construct a first passage time probability distribution for the first stage and for that we require the kernel (to which we shall also refer to as the propagator)
with \emph{exit} boundary conditions at $\phi_0$, achieved by setting $P_{\rm HJ}(\phi_0, \alpha)=0$ \cite{vanKampen}. The probability current at that point then injects probability for the second stage of the diffusion - one can thus think of the HJ branch as a ``conveyor belt''  feeding the second diffusive process at a single point $\phi_0$.    

The stochastic equation describing the IR field dynamics reads,  
\begin{equation} 
d \phi= -2\frac{\partial H}{\partial \phi} d \tau+\xi_\phi d\tau  
\,,
\label{phi-stochastic:2}
\end{equation} 
where  $\xi_\phi$ is the noise  generated by the flow of modes between the UV and IR sectors of the theory. When treated perturbatively, due to an effectively time-dependent cutoff, the leading order contribution occurs at the tree level~\cite{Starobinsky:1986fx} and on super-Hubble scales the noise is, to a good approximation, of Markovian type, 
\beq
\langle \xi_\phi(\tau) \xi_\phi(\tau^\prime) \rangle = \mathcal{A}\delta(\tau-\tau^\prime)
\,,
\label{noise noise correlator}
\eeq
where $\tau=\int^t N(t,\vec x)dt^\prime$ denotes a reparametrization invariant time.
The coupling between the ultraviolet and long wavelength modes can then be approximated by its tree level expression, 
\begin{equation}
\mathcal{A}= \frac{(\sigma aH)^3}{2\pi^2}(1-\epsilon)H|\phi(\tau,k)|^2_{k=\sigma aH}
\left[1+{\cal O}(\kappa^2 H^2)\right]
\,,
\label{noise amplitude}
\end{equation}
 where $\kappa^2 = 16\pi G$ is the loop counting parameter of quantum gravity
and 
$\sigma$ sets the highest (ultraviolet cutoff) energy scale of the long-wavelength theory. For example, 
when $\sigma=1$, the highest scale (smallest wavelength) is the Hubble scale, when $\sigma\ll 1$,
the highest scale is much smaller than the Hubble scale (or equivalently wavelength much longer than $H^{-1}$). Since there can be no secular enhancement in the loop corrections in~(\ref{noise amplitude}),
the loop suppression  factor, $\kappa^2 H^2\lesssim 10^{-12}$ represents a fair estimate of the accuracy of the stochastic approximation scheme developed in this work. In order to estimate the noise amplitude~(\ref{noise amplitude}),
in what follows we work in the approximation $\epsilon\approx 0$ and set $\sigma\ll 1$, in which
case $|\phi(\tau,k)|^2\simeq H_0^2/(2k^3)$ and
the noise amplitude~(\ref{noise amplitude}) simplifies to, 
\begin{equation}
\mathcal{A}\approx \frac{H_0^3}{4\pi^2} 
\,,
\label{noise amplitude: dS}
\end{equation}
which is the approximation we use below. Rigorous proof that Starobinsky's stochastic inflation \cite{Starobinsky:1986fx} reproduces 
the correct infrared dynamics on de Sitter can be found in Ref.~\cite{Tsamis:2005hd,Finelli:2008zg,Finelli:2010sh,Garbrecht:2013coa,Garbrecht:2014dca,Moss:2016uix} 
for interacting scalar field theories and in 
Ref.~\cite{Prokopec:2007ak} for quantum scalar electrodynamics. These works demonstrate that stochastic inflation not only
reproduces the leading infrared logarithms at each order in perturbation theory, 
but (when summed up) they also reveal what happens at late times in the deep nonperturbative 
regime when the large logarithms overwhelm small coupling constants, a point first made in \cite{Starobinsky:1994bd}.
To directly compute the curvature perturbation we use the number of e-folds $\alpha$ as the time variable.\footnote{We are therefore using a uniform expansion gauge in perturbation theory terminology. This is not fully equivalent to choosing spatially flat time slices but, as pointed out in \cite{Salopek:1990jq}, volume preserving shape deformations are not interesting dynamically in our approximations. See also appendix A on this point.}  This is in fact required for consistency with standard cosmological pertubation theory, see e.g.~\cite{Vennin:2015hra}. 
We therefore have the following branches of the evolution: 
\vspace{0.3cm}

\noindent \underline{HJ Branch}:
The Langevin equation on the HJ branch is
\begin{equation}
\frac{d\phi}{d\alpha} = -2\frac{\partial\ln H(\phi,\phi_0)}{\partial\phi} + \frac{H(\phi,\phi_0)}{2\pi}\xi(\alpha)
\end{equation}
with 
\begin{equation}
\langle\xi(\alpha)\xi(\alpha')\rangle = \delta\left(\alpha-\alpha'\right)
\end{equation}
and an absorbing boundary condition at $\phi=\phi_0$. $H(\phi,\phi_0)$ is the solution to the HJ equation with $\phi_0$ determined by the initial velocity of the field. When expressed as a Fokker-Planck equation this implies that the probability density $P_{\rm HJ}(\phi,\alpha)$ on the HJ branch obeys,  
\begin{eqnarray}
\frac{\partial P_{\rm HJ}}{\partial \alpha}  &=& -\frac{\partial}{\partial\phi}\! \left(\!-2\frac{\partial\ln H}{\partial\phi} P_{\rm HJ}\!\right) + \frac{1}{2}\frac{\partial^2}{\partial\phi^2}\!\left(\!\frac{H^2}{4\pi^2}P_{\rm HJ}\!\right)
\label{Fokker Planck HJ branch} 
\\
&\equiv&-\frac{\partial J}{\partial\phi}
\,, 
\label{Injected current HJ branch} 
\end{eqnarray} 
to be solved with the boundary condition $P_{\rm HJ}(\phi_0,\alpha)=0$ which implies that once a random walker $\phi$ among the ensemble ventures to $\phi=\phi_0$, it is removed - see {\it e.g.} 
Ref.~\cite{vanKampen} for a detailed discussion of this boundary condition's use in exit problems. 

\vspace{0.3 cm}

\noindent \underline{$H_0$ (de Sitter) branch}: Once the stochastically evolving field at a spatial point reaches $\phi_0$, it is removed from the HJ branch and is injected into the degenerate $V=V_0$ (de Sitter) branch, where $H=H_0=\sqrt{V_0/3}$. 
It then diffuses along the semi-infinite branch $\phi\in [\phi_{\rm e},\infty)$ of the flat potential $V=V_0$ according to the Langevin equation,
\begin{equation}
\frac{d\phi}{d\alpha} =  \frac{H_0}{2\pi}\xi(\alpha)
\end{equation}
again with
\begin{equation}
\langle\xi(\alpha)\xi(\alpha')\rangle = \delta\left(\alpha-\alpha'\right)
\,,
\end{equation}
where now the influx from the HJ branch must also be accounted for. When the exit point $\phi_{\rm e}$ of the $V=V_0$ branch is reached inflation may end, for example by entering a non-slow roll region or by instant reheating, or the field may enter a subsequent slow roll phase. In either case, the quantity of interest is the number of e-folds until $\phi_{\rm e}$ is reached which is a stochastic quantity.

We can write the probability distribution for $\phi$ on the $H_0$ branch as 
\begin{equation}\label{total prob}
P_{V_0}(\phi) = P_D(\phi) + P_{\star}\,\delta\left(\phi - \phi_0\right)
\end{equation}
where $P_D$ is that part which has diffused along the $V=V_0$ surface while $P_{\star}$ denotes the probability at $\phi_0$ leaking in from the HJ branch. Its contribution to the Fokker-Planck equation on the $V=V_0$ branch can be computed as follows: in a time interval between $\alpha$ and $\alpha + d\alpha$ the amount of random walkers flowing in from the HJ branch is 
\begin{equation}
dP_\star = \int_{\phi_0}^\infty \, \big[ P_{\rm HJ}\left(\phi,\alpha+d\alpha\right) 
\!-\! P_{\rm HJ}\left(\phi,\alpha\right)\big]d\phi
\end{equation} 
Therefore, 
\begin{equation}\label{Pstar evol}
\frac{\partial P_\star}{\partial \alpha} 
  = \int_{\phi_0}^\infty\frac{\partial P_{\rm HJ}}{\partial \alpha} d\phi
  = J(\phi_0,\alpha)
\,.
\end{equation}
and the Fokker-Planck equation for $P_{V_0}$ can then be written as 
\begin{equation}\label{V0 branch FP}
\frac{\partial P_{V_0}}{\partial \alpha}  =  \frac{1}{2}\frac{\partial^2}{\partial\phi^2}\left(\frac{V_0}{12\pi^2}P_{V_0}\right) + J(\phi_0,\alpha)\delta\left(\phi-\phi_0\right)
\,,
\end{equation}     
where, recalling that $P_{\rm HJ}(\phi_0) = 0$ and $\partial_\phi H(\phi_0)=0$,
\begin{equation}
J(\phi_0,\alpha)=\frac{V_0}{24\pi^2}\frac{\partial P_{\rm HJ}}{\partial\phi} \Big|_{\phi_0}
\,.
\label{injection current}
\end{equation}

The probability distribution for the number of e-folds it takes for the field to reach $\phi_{\rm e}$ can be obtained from knowledge of $P_{D}$ by noting that once the random walker has been injected into the $V_0$ branch and has started diffusing, the probability it hasn't yet crossed $\phi_{\rm e}$ by the time of $\mathcal{N}$ e-folds is the same as that of inflation lasting longer than $\mathcal{N}$ e-folds: 
\begin{eqnarray}\label{P-rho integral relation}
\text{Prob}(\text{Inflationary duration}>\mathcal{N})&=&\int\limits_\mathcal{N}^\infty\varrho(\alpha) d\alpha\nonumber \\
=\int\limits_{\phi_{\rm e}}^\infty P_{D}(\phi,\mathcal{N}) \,d\phi &&
\end{eqnarray}
where we denoted the probability that inflation lasts (more precisely $\phi_{\rm e}$ is reached) between $\alpha$ and $\alpha + d\alpha$ e-folds by $\varrho(\alpha)$.
Therefore
\begin{equation}
\varrho(\mathcal{N}) = -\frac{\partial}{\partial\mathcal{N}}\int\limits_{\phi_{\rm e}}^\infty  P_{D}(\phi,\mathcal{N}) \,d\phi\,.
\label{exit probability}
\end{equation}
Using (\ref{total prob}), (\ref{Pstar evol}) and (\ref{V0 branch FP}) we obtain simply 
\begin{equation}\label{efold probability}
\varrho(\mathcal{N}) = \frac{V_0}{24\pi^2}\frac{\partial P_{V_0}(\phi,\mathcal{N})}{\partial\phi} \Big|_{\phi_{\rm e}}\,.
\end{equation} 


\subsection{Computing $P_{\rm HJ}$}
\label{Computing PHJ}

In order to obtain the current flowing into the $V_0$ branch from (\ref{injection current}) we first need to compute $P_{\rm HJ}$, the probability distribution on the HJ branch. To obtain simple analytic expressions, we will make the approximation that $\phi$ is close to $\phi_0$ on the HJ branch, corresponding to a small initial velocity. This is justified since  
\begin{equation}
\phi_{\rm in}-\phi_0 = {\rm arc}\sinh \left(-\frac{\Pi_{\rm in}}{\sqrt{2}V_0^{1/2}}\right)\simeq -\frac{\Pi_{\rm in}}{\sqrt{2}V_0^{1/2}}\ll1 
\end{equation}
assuming that the field enters the USR regime from a previous slow roll phase. We will tackle the more general problem in an upcoming publication \cite{Prokopec:2019ii}. We therefore take the HJ branch stochastic dynamics to be (see~\ref{general solution of HJ: USR2})
\begin{equation}
\frac{d\phi}{d\alpha} \simeq  -3\left(\phi\!-\!\phi_0\right) + \frac{H_0}{2\pi}\xi(\alpha)
\label{PHJ: linearized}
\end{equation}
with an exit boundary conditions at $\phi_0$. Setting $\chi =\frac{\sqrt{12}\pi}{H_0}\left(\phi-\phi_0\right)$ the corresponding Fokker-Planck equation~(\ref{Fokker Planck HJ branch}) for
the probability density $P_{\rm HJ}(\chi,\alpha)$ reads,
\begin{equation}
\frac{\partial P_{\rm HJ}}{\partial \alpha} 
 \simeq 3\frac{\partial}{\partial \chi} \left(\chi P_{\rm HJ}\right) 
     + \frac{3}{2}\frac{\partial^2 P_{\rm HJ}}{\partial \chi^2}
\,.
\end{equation} 
Writing 
\begin{equation}
P_{\rm HJ}(\chi,\alpha) = C e^{\frac32\alpha-\frac{1}{2}\chi^2}
   \Psi(\chi,\alpha)\,,
\label{rescaling P to psi}
\end{equation} 
where $C$ is a constant independent of $\alpha$ and $\chi$ but dependent on 
the choice of the initial state, $\Psi(\chi,\alpha)$ obeys
\begin{equation}
-\frac13\frac{\partial \Psi}{\partial \alpha} 
   = \frac12\left(-\frac{\partial^2}{\partial \chi^2} + \chi^2\right)\Psi
\,,
\label{EOM psi}
\end{equation}
and the problem reduces to the quantum mechanical kernel for the {\it simple harmonic oscillator} (SHO),
with a mass $m$ and frequency $\omega$ given by $m\omega\rightarrow \hbar H_0^2/(12\pi^2)$) 
in imaginary time $\alpha=i\omega t/(3\hbar)$
(or $t=-3i\hbar\alpha/\omega$),
see {\it e.g.}~\cite{Sakurai:2011zz}. The free propagator, also known in the literature on stochastic processes as the {\it Mehler heat kernel}~\cite{Pauli:2000}, is given by 
\begin{widetext}
	\begin{equation}
K_M(\chi,\alpha;\chi_{\rm in},\alpha_{\rm in}) = \frac{1}{\sqrt{2\pi\sinh[3(\alpha\!-\!\alpha_{\rm in})]}}
\exp\left(-\frac{\coth[3(\alpha\!-\!\alpha_{\rm in})](\chi^2\!+\!{\chi_{\rm in}}^2)}{2}
 +\frac{\chi\chi_{\rm in}}{\sinh[3(\alpha\!-\!\alpha_{\rm in})]}\right)
 \label{free kernel HJ branch}
\end{equation}
\end{widetext}
which for small time intervals tends to
\begin{equation}
\lim\limits_{\alpha\rightarrow \alpha_{\rm in}}K_M(\chi,\alpha;\chi_{\rm in},\alpha_{\rm in})=\delta(\chi\!-\!\chi_{\rm in})
\,.
\end{equation}

Since a random walker is `removed' upon reaching $\chi=0$ ($\phi=\phi_0$), 
for the problem at hand we do not require the free, but rather the absorbtive kernel. 
Due to the symmetry of the effective potential in which the dynamics takes place, it can be obtained from the full kernel (\ref{free kernel HJ branch})
by adding to it a free \emph{mirror} kernel at $-\chi$, giving
\begin{eqnarray}
 \Psi(\chi,\alpha) &=&  K_M(\chi,\alpha;\chi_{\rm in},\alpha_{\rm in}) - K_M(-\chi,\alpha;\chi_{\rm in},\alpha_{\rm in}) \nonumber \\
 &=&\sqrt{\frac{2}{\pi}}\frac{\exp\left(-\frac12\coth[3(\alpha\!-\!\alpha_{\rm in})](\chi^2\!+\!{\chi_{\rm in}}^2)\right)}{\sqrt{\sinh[3(\alpha\!-\!\alpha_{\rm in})]}}    \nonumber\\
 &&\times \quad \sinh\left(\frac{\chi\chi_{\rm in}}{\sinh[3(\alpha\!-\!\alpha_{\rm in})]}\right)   
 \label{wave function HJ branch}
\end{eqnarray}
which ensures the correct boundary condition is satisfied. 
The properly normalized $P_{\rm HJ}$ is then obtained from~(\ref{rescaling P to psi}),
\begin{equation}
P_{\rm HJ}(\varphi,\alpha) 
=\frac{6\pi}{\sqrt{V_0}} 
   \exp\left(\frac32(\alpha\!-\!\alpha_{\rm in})-\frac{1}{2}(\chi^2\!-\!\chi_{\rm in}^2)\right)
   \Psi(\chi,\alpha) 
\,,
\label{rescaling psi back to P}
\end{equation} 
where $C$ in Eq.~(\ref{rescaling P to psi}) is chosen such 
that in the limit $\alpha\rightarrow \alpha_{\rm in}$ reduces 
to~\footnote{
Strictly speaking, in the limit $\alpha\rightarrow\alpha_{\rm in}$
 the probability density~(\ref{rescaling psi back to P}) reduces  to 
$\delta(\chi\!-\!\chi_{\rm in})+\delta(\chi\!+\!\chi_{\rm in})$. Since 
the domain of validity of~(\ref{wave function HJ branch}) is the HJ branch on which $\chi\geq 0$, 
the second delta function is discarded. 
}
\begin{equation}
P_{\rm HJ}(\varphi,\alpha\rightarrow\alpha_{\rm in}) 
      =\frac{6\pi}{\sqrt{V_0}}\delta(\chi\!-\!\chi_{\rm in})=\delta(\phi\!-\!\phi_{\rm in})\,,
\label{initial condition for absobtive kernel HJ branch}
\end{equation}
for $(\chi\geq 0, \chi_{\rm in}>0)$. A general probability distribution on the HJ branch can be obtained by convolving
the absorbtive kernel~(\ref{wave function HJ branch}) 
with the initial probability distribution.

\begin{figure}[t]
\vskip -0.8cm\label{fig:current}
	\begin{center}
		\includegraphics[scale=0.5]{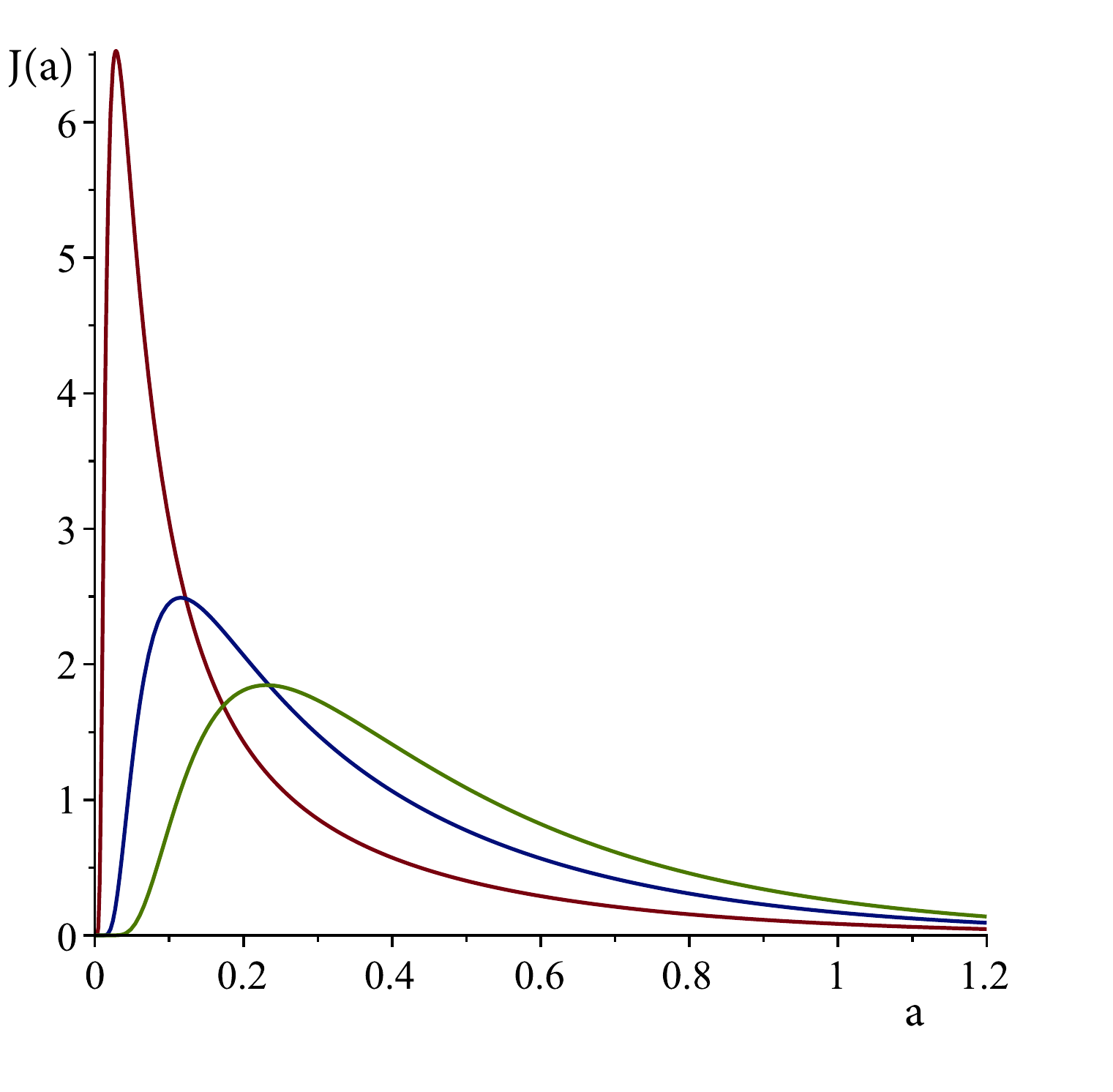}
\vskip -0.6cm
		\caption{The current $J(\alpha)$ injected from the HJ branch into the flat branch
		at $\phi=\phi_0$ for $\chi_{\rm in}=0.5,1,1.5$ (from leftmost to rightmost curve).}
	\end{center}	
\vskip -0.3cm
\end{figure}

The injected current~(\ref{injection current}), 
$J = [\sqrt{V_0}/(4\pi)]\partial_\chi P_{\rm HJ}|_{\chi\rightarrow 0}$,
 into the flat $V_0$ branch
is obtained by taking a derivative of~(\ref{rescaling psi back to P}),
\begin{eqnarray}
J(\alpha)&=&
   \frac{6\pi}
            {\big[2\pi\sinh[3\Delta\alpha]\big]^{3/2}}\nonumber\\
&\times& \exp\left[\frac32\Delta\alpha
          \!-\!\frac12\big(\coth[3\Delta\alpha]\!-\!1\big)\chi_{\rm in}^2\right]\chi_{\rm in}
\,,
\label{injection current:2}
\end{eqnarray}
which rises at early times $\Delta\alpha=\alpha\!-\!\alpha_{\rm in}\ll 1$ as,
\begin{equation}
J(\alpha)|_{\Delta \alpha\ll 1} = \frac{\chi_{\rm in}}{\sqrt{6\pi}(\Delta \alpha)^{\frac{3}{2}}}   e^{-\frac{\chi_{\rm in}^2}{6\Delta\alpha}}
\,,
\label{injection current:early times}
\end{equation}
whereas at late times, when $\Delta\alpha=\alpha\!-\!\alpha_{\rm in}\gg 1$,
it decays exponentially, 
\begin{eqnarray}
J(\alpha)|_{\Delta \alpha\gg 1}
=\frac{6}{\sqrt{\pi}}\chi_{\rm in}
e^{-3\Delta\alpha}  +\mathcal{O}\left(e^{-9\Delta\alpha}\right)
\,.
\label{injection current:late times}
\end{eqnarray}
The current $J(\alpha)$ for three different values of $\chi_{\rm in}$ can be seen in figure \ref{fig:current}.
The current increases from zero at $t=0$ ($\alpha=\alpha_{\rm in}$), 
peaks and then decays exponentially as $\propto e^{-3(\alpha\!-\!\alpha_{\rm in})}$,
 see~(\ref{injection current:early times})~and~(\ref{injection current:late times}).   
%


\subsection{Computing $P_{\rm V_0}$}
\label{Computing PV0}

Assuming that the initial field distribution lies entirely at the HJ branch, equation (\ref{V0 branch FP}) must be supplemented by the initial condition $P(\chi,0) = 0$ and the solution can therefore be written as
\begin{equation}
P_{V_0}(\phi,\alpha) = \int_{\alpha_{\rm in}}^\alpha du \,G_{\rm e}(\phi-\phi_0, \alpha-u)J(u) 
\,,
\label{sourced probability on flat branch}
\end{equation}
where $G_{\rm e}\left(\phi-\phi',\alpha-\alpha'\right)$ is the diffusive Green function with exit boundary condition at 
$\phi_{\rm e}$, also known as the absorptive kernel. 
%
As above, it is straightforwardly constructed from the well known unrestricted diffusive kernel along an infinite interval
\begin{equation}
G\left(\phi,\phi',\Delta\alpha\right)=\sqrt{\frac{2\pi}{H_0^2\Delta\alpha}}\exp\!\left(\!-\frac{2\pi^2}{H_0^2\Delta\alpha}(\phi\!-\!\phi')^2\right)
\end{equation}
where $\Delta\alpha = \alpha\!-\!\alpha'$, by subtracting the same kernel but with $\phi'$ reflected on $\phi_{\rm e}$: $\phi\rightarrow 2\phi_{\rm e}-\phi$, giving
\begin{eqnarray}
G_{\rm e}\left(\phi,\phi',\Delta\alpha\right)&=&\sqrt{\frac{2\pi}{H_0^2\Delta\alpha}}
\Bigg[\exp\!\left(\!-\frac{2\pi^2}{H_0^2\Delta\alpha}(\phi\!-\!\phi')^2\right)\nonumber\\
&-&\exp\!\left(\!-\frac{2\pi^2}{H_0^2\Delta\alpha}(2\phi_{\rm e}\!-\phi\!-\!\phi')^2\right)
\Bigg]\,.
\label{absorptive kernel}
\end{eqnarray} 
This imposes the correct boundary conditions, $G_{\rm e}\left(\phi_{\rm e},\phi',\Delta\alpha\right)=0$
and $G_{\rm e}\left(\phi,\phi',\Delta\alpha\rightarrow 0\right)=\delta(\phi\!-\!\phi')$ for $\phi,\phi'\geq \phi_{\rm e}$. 
The limits of integration 
in~(\ref{sourced probability on flat branch}) are determined by imposing that no current can be sourced before the beginning of inflation at $\alpha_{\rm in}$
(lower limit) and that no current can be sourced in the future of $\alpha$
(upper limit).

To compute the probability distribution for the field on the flat branch we use the convolution integral~(\ref{sourced probability on flat branch})
with the absorbtive kernel 
is~(\ref{absorptive kernel}) and the injected current~(\ref{injection current:2}) to obtain,
\begin{widetext}
\begin{eqnarray}
P_{V_0}(\phi,\alpha) 
 =\frac{3\chi_{\rm in}}{H_0} \int_{0}^\alpha \frac{du}{\sqrt{\alpha\!-\!u}}
\left[\exp\!\left(\!-\frac{\chi^2}{6(\alpha\!-\!u)}\right)
\!-\!\exp\!\left(\!-\frac{(\chi\!-\!2\chi_e)^2}{6(\alpha\!-\!u)}\right)
\right] \frac{\exp\left[\frac32u
	\!-\!\frac12\big(\coth(3u)\!-\!1\big)\chi_{\rm in}^2\right]}
            {\big[\sinh(3u)\big]^{3/2}}
\,,
\label{probability on flat branch:2}
\end{eqnarray}
\end{widetext}
where
$\chi =\sqrt{12}\pi(\phi-\phi_0)/H_0$, $\chi_e =\sqrt{12}\pi(\phi_{\rm e}\!-\!\phi_0)/H_0$ and we set, for simplicity, 
$\alpha_{\rm in}=0$~\footnote{One can always recover the dependence on $\alpha_{\rm in}$
by noting that the integral~(\ref{probability on flat branch:2})
 is a function of $\alpha\!-\!\alpha_{\rm in}$}. 

\subsection{Probability density for the e-fold number}
 
From $P_{\rm V_0}$ we can directly compute the e-fold probability density using (\ref{efold probability})
\begin{eqnarray}\label{e-folds-prob-int}
\!\!\!\!\!&\varrho&\!\!(\mathcal{N}) = \frac{\sqrt{3}}{2\pi}\left(-\chi_{\rm e}\chi_{\rm in}\right) \nonumber \\
\!\!\!\!&\times&\!\!\!\int_{0}^\mathcal{N} \!\!\!\!du\,\frac{\exp\left[-\frac{\chi_{\rm e}^2}{6\left(\mathcal{N}-u\right)}+\frac32u
	\!-\!\frac12\big(\coth(3u)\!-\!1\big)\chi_{\rm in}^2\right]}
{\big[\left(\mathcal{N}-u\right)\sinh(3u)\big]^{3/2}}
\end{eqnarray}
where we note that $\chi_{\rm e}<0$ by definition. Although the above integrals cannot be evaluated analytically, an approximate evaluation of (\ref{probability on flat branch:2}) can be performed by noting that the dominant 
dependence on $u$ sits in the exponent and the integral can be well approximated 
by a steepest descent method presented in Appendix~B. There is a very simple case, namely if the integral is dominated by $u\ll 1$
and if $\alpha\gg 1$, then it evaluates to, 
\begin{eqnarray}
P_{V_0}(\!\!&\phi&\!,\alpha) 
 \approx\frac{\sqrt{12}\pi}{H_0}\frac{e^{-\chi_{\rm in}^2/2}}{\sqrt{6\pi\Delta\alpha}}\nonumber\\ 
 &\times& \!\left[\exp\left(-\frac{\chi^2}{6\Delta\alpha}\right)
  -\exp\left(-\frac{(\chi\!-\!2\chi_e)^2}{6\Delta\alpha}\right)\right]
\,,
\label{probability on flat branch:simple}
\end{eqnarray}
which is, up to the factor $e^{-\chi_{\rm in}^2/2}$, equal to the absorbtive kernel 
$G_e(\phi\!-\!\phi_0;\Delta\alpha)$
($\Delta\alpha=\alpha\!-\!\alpha_{\rm in}$) in Eq.~(\ref{absorptive kernel}). Therefore 
\beq\label{eq:asymptotic-N3/2-efolds}
\varrho(\mathcal{N})\approx \sqrt{\frac{6}{\pi}}e^{-\frac{\chi_{\rm in}^2}{2}} \frac{e^{-\frac{\chi_{\rm e}^2}{6\mathcal{N}}}}{\mathcal{N}^{3/2}}
\eeq

\begin{figure}[t]
\vskip -0.8cm
	\begin{center}
		\includegraphics[scale=0.47]{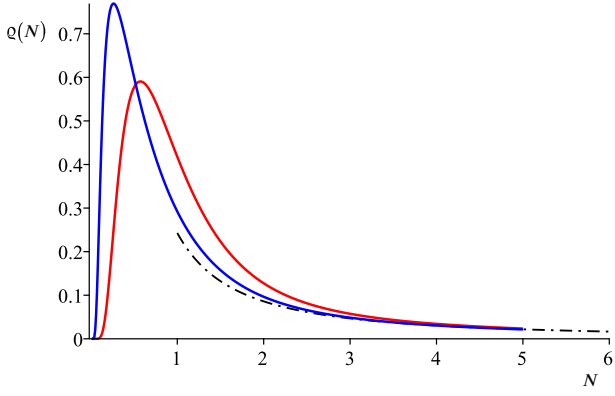}
\vskip -0.5cm
		\caption{The probability distribution $\varrho(\mathcal{N})$ defined by (\ref{e-folds-prob-int}) for $\chi_{\rm e}=-1$ and $\chi_{\rm in}=0.5$ (leftmost, blue curve) or $\chi_{\rm in}=1.5$ (rightmost, red curve). The dashed-dotted line indicates the asymptotic $\varrho\propto\mathcal{N}^{-3/2}$ behavior for large $\mathcal{N}$, see (\ref{eq:asymptotic-N3/2-efolds}). This deep non-Gaussian tail is responsible for eternal inflation.}
	\end{center}	
\label{fig:rho1}
\vskip -0.3cm
\end{figure} 

We see that although $\varrho(\mathcal{N})$ is normalisable, the probability distribution does not decay fast enough as $\phi\rightarrow \infty$ and therefore all moments are infinite: $\langle \mathcal{N}^n \rangle = \infty$ for $n\geq 1$. A numerical evaluation of the probability density $\varrho(\mathcal{N})$ is plotted in figure~5.2. 
For large $\mathcal{N}$ the distribution tends to an $\propto\mathcal{N}^{-3/2}$
 decay, which is in agreement with our analytic estimate. This  reflects the fact that if the precipice signified by $\phi_{\rm e}$ is beyond $\phi_0$, the field settles into free diffusion along the half-line towards $\phi \rightarrow \infty$, a situation termed infinite inflation in \cite{Assadullahi:2016gkk, Vennin:2016wnk}.  

The endless diffusion towards $\phi\rightarrow \infty$ would of course not occur if the field was injected into the de Sitter branch from a prior slow-roll regime. We will deal with this in more detail in \cite{Prokopec:2019ii} where more complete models are studied. A simple way to regulate this infinite diffusion would be to erect a reflecting wall at, or close to $\phi_{\rm in}$. In \cite{Pattison:2017mbe} this is shown to indeed lead to a distribution with finite moments and hence a finite curvature perturbation.

\section{The case $\phi_{\rm e}>\phi_0$: USR with graceful exit}
\label{graceful exit}

\begin{figure}[t]
\vskip -0.5cm
	\begin{center}
		{\includegraphics[scale=0.43]{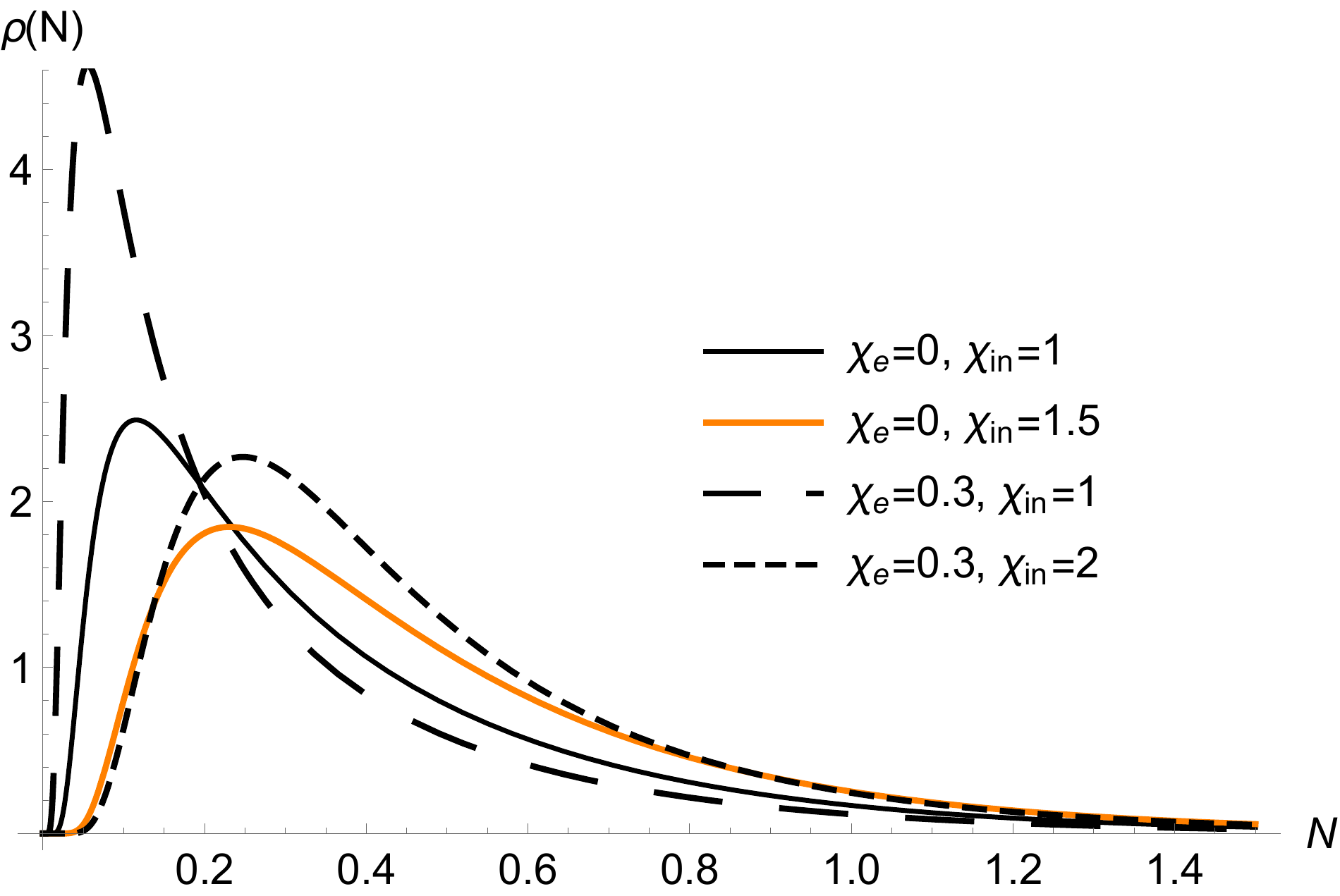}}
		\caption{The probability distribution~(\ref{exit probability USR}--\ref{exit probability USR:5})
		in the USR model with graceful exit as a function of the number of e-folds 
		$\Delta\mathcal{N}=\mathcal{N}-\mathcal{N}_{\rm in}$ with $\mathcal{N}_{\rm in}=0$
 and given in Eq.~(\ref{exit probability USR:5}) for four choices of $(\chi_{\rm e},\chi_{\rm in})$:
		(0, 1) (solid black); (0, 1.5) (solid orange); (0.3, 1) (long dashes) and (0.3, 2) (short dashes).
		We see that for larger $\chi_{\rm e}$ ($\chi_{\rm in}$) inflation gets shorter (longer), 
		which is what as one would expect.
}
		\label{Figure probability graceful}
	\end{center} 
	\vskip -0.5cm
\end{figure}

We saw in the previous section that if the initial velocity of the field does not suffice to carry it beyond $\phi_{\rm e}$ ($\phi_{\rm e}<\phi_0$), eternal inflation sets in on the semi-line $[\phi_{\rm e},\infty)$ and the curvature perturbation is infinite, as signified by the divergence of all moments of 
$\mathcal{N}$. We show in this section that this is not true when the exit point $\phi_{\rm e}$ occurs on the HJ branch, {\it i.e.}~before the asymptotic point $\phi=\phi_0$ at which the classical trajectory of $\phi$ would terminate. In other words we now assume that,
\begin{equation}
\phi_{\rm e}\geq \phi_0\,,
\label{URD with exit}
\end{equation}
and show that this model of inflation exhibits a graceful exit. The probability density $P_{HJ}(\phi,\alpha)$ is 
then of the form~(\ref{rescaling psi back to P}), but with $\Psi(\chi,\alpha)$ given by 
the absorbtive kernel mirrored at $\chi_{\rm e}$, see Eq.~(\ref{wave function HJ branch}),
\begin{widetext}
	\begin{eqnarray}
\Psi(\chi,\alpha) &=&  K_M(\chi,\alpha;\chi_{\rm in},\alpha_{\rm in}) - K_M(2\chi_e-\chi,\alpha;\chi_{\rm in},\alpha_{\rm in}) \nonumber \\
&=&\frac{1}{\sqrt{2\pi\sinh[3(\alpha\!-\!\alpha_{\rm in})]}}\Bigg\{
\exp\left(-\frac12\coth[3(\alpha\!-\!\alpha_{\rm in})](\chi^2\!+\!{\chi_{\rm in}}^2)
\!+\!\frac{\chi\chi_{\rm in}}{\sinh[3(\alpha\!-\!\alpha_{\rm in})]}\right)
\nonumber\\
&& - \,
\exp\left(-\frac12\coth[3(\alpha\!-\!\alpha_{\rm in})]\left[(2\chi_{\rm e}\!-\!\chi)^2\!+\!{\chi_{\rm in}}^2\right]
\!+\!\frac{(2\chi_{\rm e}\!-\!\chi)\chi_{\rm in}}{\sinh[3(\alpha\!-\!\alpha_{\rm in})]}\right)\Bigg\}
,\quad
\label{wave function HJ branch: graceful exit}
\end{eqnarray}

The probability $\varrho(\mathcal{N})$
in~(\ref{exit probability})
that inflation ends in the interval $[\mathcal{N},\mathcal{N}+d\mathcal{N})$ of e-folds is then,
\begin{eqnarray}
\varrho(\mathcal{N})
&=& -\frac{\partial}{\partial\mathcal{N}}\int_{\phi_{\rm e}}^\infty d\phi P_{\rm HJ}(\mathcal{N},\phi)
\label{exit probability USR}\\
&&\hskip -1.5cm
=- \frac{\partial}{\partial\mathcal{N}}\biggr\{
\frac12{\rm erfc}\Big[\sqrt{1\!+\!n}\Big(\chi_{\rm e}\!-\!e^{-3\Delta \mathcal{N}}\chi_{\rm in}\Big)\Big]
\label{exit probability USR:2}
-\,{\rm exp}\Big[\chi_{\rm in}^2\!-\!\chi_{\rm e}^2
\!-\!(\chi_{\rm in}\!-\!e^{-3\Delta \mathcal{N}}\chi_{\rm e})^2\Big]
\frac12 {\rm erfc}\Big[\sqrt{n}\left(\chi_{\rm in}\!-\!e^{-3\Delta \mathcal{N}}\chi_{\rm e}\right)\Big]
\biggr\}\quad
\label{exit probability USR:3}\\
&&\hskip -1.5cm 
= \frac{3\sqrt{n}}{\sqrt{\pi}}
\Bigl[ 2(1\!+\!n)\chi_{\rm in}\!-\! (3\!+\!2n)e^{-3\Delta \mathcal{N}}\chi_{\rm e}\Bigr]
{\rm exp} \Bigl[-(1\!+\!n)\left(\chi_{\rm e}\!-\!e^{-3\Delta \mathcal{N}}\chi_{\rm in}\right)^2\Bigr]
\qquad
\label{exit probability USR:4}\nonumber\\
&&\hskip -1.cm 
-3e^{-3\Delta \mathcal{N}}\chi_{\rm e}\left[\chi_{\rm in}\!-\!e^{-3\Delta \mathcal{N}}\chi_{\rm e}\right]
{\rm e}^{(\chi_{\rm in}^2-\chi_{\rm e}^2)
	-\left(\chi_{\rm in}\!-\!e^{-3\Delta \mathcal{N}}\chi_{\rm e}\right)^2}
{\rm erfc}\Big[\sqrt{n}\left(\chi_{\rm in}\!-\!e^{-3\Delta \mathcal{N}}\chi_{\rm e}\right)\Big]
\,,\quad
\label{exit probability USR:5}
\end{eqnarray}
\end{widetext}
where $n=1/(e^{6\Delta\mathcal{N}}\!-\!1)$, $\Delta\mathcal{N}=\mathcal{N}\!-\mathcal{N}_{\rm in}$
and ${\rm erfc}(z)=1\!-\!{\rm erf}(z)=\frac{2}{\sqrt{\pi}}\int_z^\infty e^{-t^2}dt$ is the complementary error function. The distribution 
$\varrho(\mathcal{N})$ in~(\ref{exit probability USR}--\ref{exit probability USR:5})
is plotted in figure~\ref{Figure probability graceful} for a few selected 
values of $\chi_{\rm e}$ and $\chi_{\rm in}$. The distribution is again strongly non-Gaussian, 
however at large $\mathcal{N}$ it falls-off exponentially as $\propto {\rm e}^{-3\mathcal{N}}$, 
such that the moments of the curvature perturbation are all finite, implying that inflation terminates.
 The first term after the curly bracket in~(\ref{exit probability USR:2})
is the standard result for the probability that the particle is located anywhere at 
$\chi>\chi_{\rm e}$, and it approaches {\it one} when 
$\chi_{\rm e}\rightarrow -\infty$, as it should, while the second term 
reduces the probability due to the absorbtive boundary condition at $\phi=\phi_{\rm e}$, 
where inflation ends. 

Eqs.~(\ref{exit probability USR}--\ref{exit probability USR:5}) contain a complete information 
for the probability distribution 
of the number of e-folds in this simple model 
(where we assumed a small initial momentum, which 
allowed us to linearize in $\phi-\phi_0$ in~(\ref{PHJ: linearized})).
To get a better understanding of $\rho(\mathcal{N})$ in~(\ref{exit probability USR}),
we shall now calculate the first few moments of the number of e-folds,
\begin{equation}
\langle {\cal N}^k\rangle 
= \int_{\rm \alpha_{\rm in}}^\infty \mathcal{N}^k  \varrho(\mathcal{N})
\,d\mathcal{N}
\,,\qquad (k=0,1,2,\cdots)
\,.
\label{N on n}
\end{equation}
Let us first look at the zeroth moment, 
\begin{equation}
\langle 1\rangle = \frac{1}{2}\left[1 + {\rm erf}[\chi_{\rm e}] + {\rm e}^{-\chi_{\rm e}^2}\right]
\qquad (\chi_{\rm e}\geq 0)
\,.
\label{0th moment of N}
\end{equation}
When $0\leq \chi_{\rm e}\lesssim 1$  
this is, as one would expect, of the order one. 
One can account for the fact that~(\ref{0th moment of N}) is not exactly equal to one 
by dividing  $\langle {\cal N}^n\rangle$ by $\langle 1\rangle$.
The moments of ${\cal N}$ are considerably more difficult to calculate, and therefore in what 
follows for simplicity we consider the case, $\chi_{\rm e}=0$ ($\phi_{\rm e}=\phi_0$).
Then the probability distribution~(\ref{exit probability USR}--\ref{exit probability USR:5}) reduces to,
\begin{equation}
\varrho(\mathcal{N}) d\mathcal{N}
=\frac{6\sqrt{n}}{\sqrt{\pi}}(1\!+\!n)\chi_{\rm in}{\rm e}^{-n\chi_{\rm in}^2}\,d\mathcal{N}
\,.
\label{exit probability USR:simple}
\end{equation}
It pays off to convert this into the probability per unit $dn=-6n(1+n)d\mathcal{N}$,
\begin{equation}
G(n)dn
=\frac{\chi_{\rm in}}{\sqrt{\pi n}}{\rm e}^{-n\chi_{\rm in}^2}dn
\,,
\label{exit probability USR: even simpler}
\end{equation}
such that the $k$-th moment in~(\ref{N on n}) gives, 
\begin{equation}
\langle {\cal N}^k\rangle 
= \frac{2}{\sqrt{\pi}}
\int_{0}^\infty dy\, {\rm e}^{-y^2}
\left[\frac16\ln\left(1+\frac{\chi_{\rm in}^2}{y^2}\right)\right]^k 
\,,
\label{N on n: simple}
\end{equation}
where $(k=0,1,2,\cdots)$ and we used $y=\sqrt{n}\chi_{\rm in}$, assuming that $\alpha_{\rm in}=0$.
Furthermore, it is useful to calculate how the number of e-folds fluctuates around its mean value,
$\langle{\cal N}\rangle$,
\begin{equation}
\langle (\Delta{\cal N})^n\rangle = \sum_{k=0}^n(-1)^k {n\choose k}
\langle {\cal N}^k\rangle\langle {\cal N}\rangle^{n-k}
\,.
\label{Delta n to nth}
\end{equation}
where $ {n\choose k}=n!/[k!(n-k)!]$ is the binomial coefficient.

The first moment in~(\ref{Delta n to nth}) can be expressed in terms of a 
generalized hypergeometric function, 
\begin{eqnarray}
\langle {\cal N}\rangle 
&=& \frac{\pi}{6}{\rm erf}(i\chi_{\rm in})\!-\! \frac{\chi_{\rm in}^2}{3}\times_{2}\!F_2\left(
\Big\{1,1\Big\},\Big\{\frac32,2\Big\},\chi_{\rm in}^2
\right)
\,.
\label{1st moment of N}
\end{eqnarray}
The higher moments are harder to evaluate analytically. Nevertheless, one can show that 
the following confluent hypergeometric function generates all the moments, 
\begin{eqnarray}
{\mathcal G}(\alpha,\chi_{\rm in})
&=&\frac{2}{\sqrt{\pi}}\int_{0}^\infty dy\, {\rm e}^{-y^2}
\left(1+\frac{\chi_{\rm in}^2}{y^2}\right)^\frac{\alpha}{6}
\nonumber\\
&=&\frac{1}{\sqrt{\pi}} \Gamma\left(\frac12-\frac{\alpha}{6}\right)
\times U\left(-\frac{\alpha}{6},\frac{1}{2},\chi_{\rm in}^2 \right)
\nonumber\\
&=&_{1}F_1\left(-\frac{\alpha}{6};\frac12;\chi_{\rm in}^2\right)
-2\frac{\Gamma\left(\frac12-\frac{\alpha}{6}\right)}{\Gamma\left(-\frac{\alpha}{6}\right)}(\chi_{\rm in}^2)^{1/2}\nonumber\\
&& \hspace{2cm} \times \, _{1}F_1\left(\frac12-\frac{\alpha}{6};\frac32;\chi_{\rm in}^2\right)
\label{generatrice of moments}
\end{eqnarray}
in the sense that 
\begin{equation}
\langle {\cal N}^k\rangle 
= \left(\frac{\partial^k}{\partial \alpha^k}{\mathcal G}(\alpha,\chi_{\rm in})\right)_{\alpha=0}
\qquad (k=1,2,3,\cdots)
\,,
\label{higher order moments}
\end{equation}
where $U$ denotes the confluent hypergeometric function. 
\begin{figure}[t]
\vskip -0.7cm
	\begin{center}
		{\includegraphics[scale=0.50]{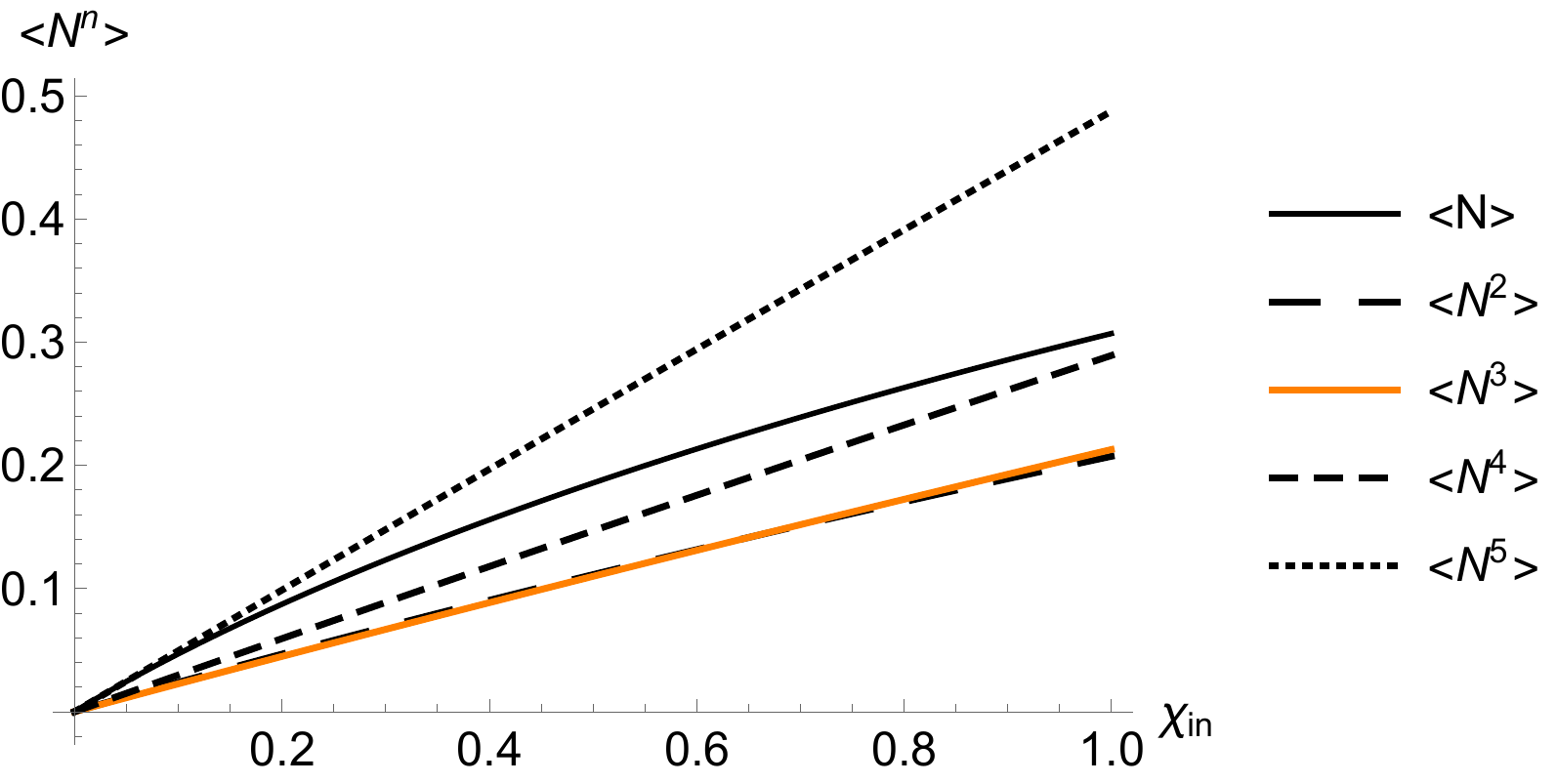} 
\vskip 0.3cm
			\includegraphics[scale=0.50]{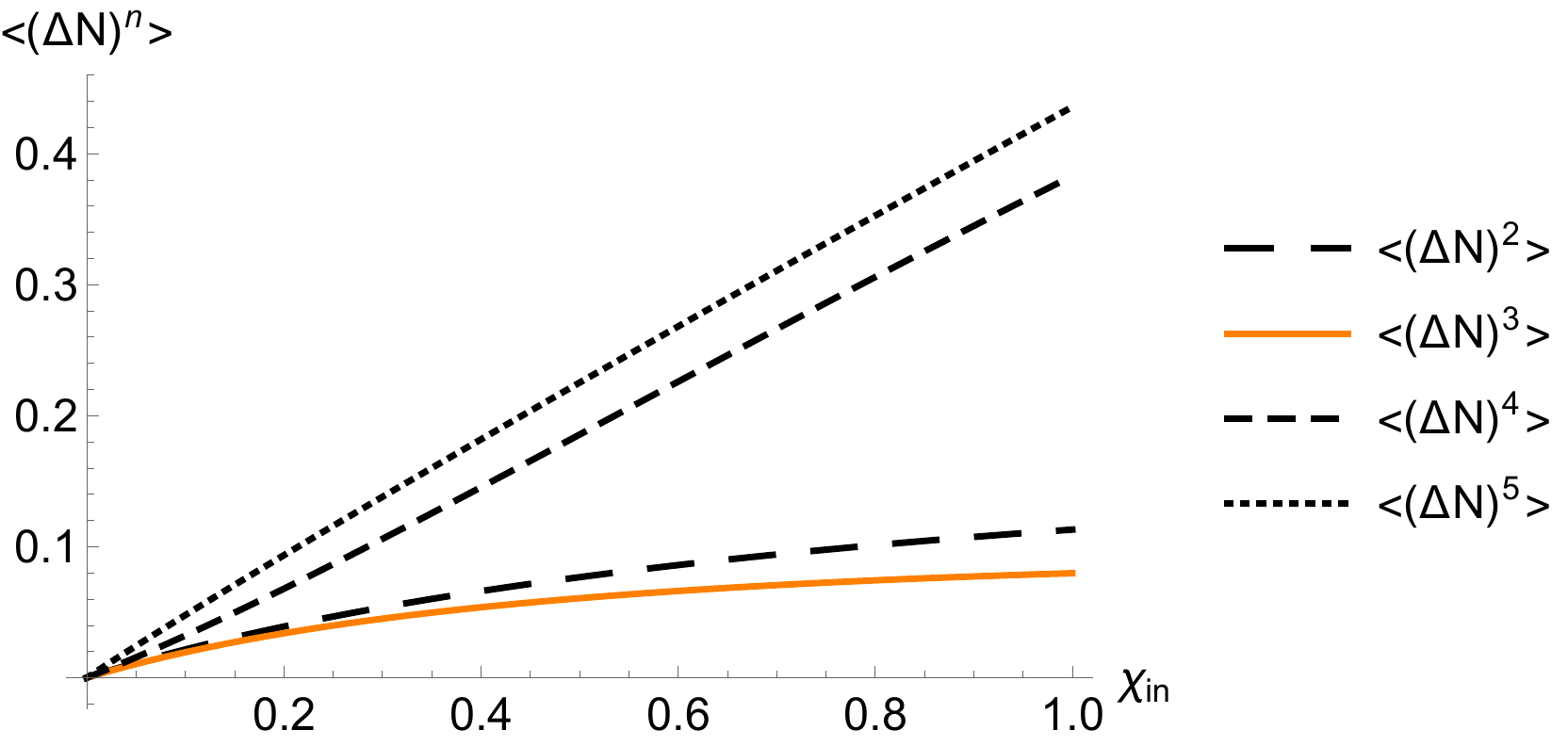}}
\vskip -0.3cm
		\caption{The first few moments of the number of e-folds for $\chi_{\rm e}=0$ 
			as a function of $\chi_{\rm in}$ in our simple USR model. We show both 
			the moments of ${\cal N}$ defined in~(\ref{N on n}) and~(\ref{N on n: simple})
			({\it upper panel}), as well as their fluctuation from the mean, 
			$\langle(\Delta {\cal N})^n\rangle$, defined in~(\ref{Delta n to nth}) 
			({\it lower panel}).   
		}
		\label{Figure moments}
	\end{center} 
	\vskip -0.5cm
\end{figure}
In figure~\ref{Figure moments} we show the first few moments
in Eq.~(\ref{N on n: simple}) and their fluctuations around the mean 
$\langle\mathcal{N}\rangle$ defined in Eq.~(\ref{Delta n to nth}). For simplicity
we choose $\chi_{\rm e}=0$ and plot our results as a function of $\chi_{\rm in}$. 
We see that the distribution is highly non-Gaussian, which is one one of the main results of this work. Because we have calculated $G(\alpha)$ with the assumption of small $\phi\!-\!\phi_0$, 
in figure~\ref{Figure moments} we plot the results only for $\chi_{\rm in}<1$. The principal conclusion is 
that the non-Gaussianities produced in USR are quite large and grow 
with $\chi_{\rm in}$, or the length of the USR supporting potential segment. On the other hand, from figure~\ref{Figure probability graceful} we see that a larger $\chi_{\rm in}$ implies a larger average number of e-folds of USR $\langle\mathcal{N}\rangle$, from which we conclude that a longer USR phase generates larger non-Gaussianities. This observation can be of crucial importance
for the generation of primordial black holes. These results are in broad agreement with the findings of \cite{Pattison:2017mbe} and it would be interesting to make a more quantitative comparison, recalling that we have fully and consistently included the gravitational effects of the field's velocity.


\section{Conclusion and discussion}
\label{Conclusion and discussion}

In this paper we established a consistent formalism for describing the quantum evolution of the large scale curvature perturbation, generated during inflation on very flat portions of the potential $V(\phi)$, fully taking into account the scalar gravitational back-reaction and the finite classical velocity for the field. This was achieved by combining a long wavelength approximation to the Einstein equations with the stochastic picture of inflationary quantum fluctuations. We found that the $0i$ Einstein equation, usually neglected in the widely used ``separate universe'' approach, leads to a single stochastic equation for the scalar field but not its velocity, the latter being fully determined by the former even beyond slow roll through a unique solution to the Hamilton-Jacobi equation (\ref{Hamilton Jacobi}). 

We then focused on a completely level potential $V=V_0$ where inflation occurs in 
an ultra slow roll (USR) regime. We assumed that $\phi\in [\phi_{\rm e},+\infty)$ and that the field is injected with some finite velocity $\Pi_{\rm in}$ at $\phi_{\rm in}$. We showed that on large (super-Hubble) scales USR is a phase space attractor, in the sense that gravitational constraints fully fix the field velocity in terms of the field $\phi$, up to a global constant $\phi_0$ determined by the initial velocity and marking the end point towards which the classical field evolution asymptotes. This is accurate up to small, exponentially decaying gradient corrections, which are highly suppressed, and thus completely irrelevant on very large scales. The value $\phi_{\rm e}$ demarcated an exit point where inflation either ends or the field enters into another region of the potential, presumably one supporting slow roll. The stochastic number of e-folds required to reach 
$\phi_{\rm e}$ directly gives the curvature perturbation. 

The inflaton dynamics 
depends crucially on the distance between the entry and exit points $|\phi_{\rm in}-\phi_{\rm e}|$ and on 
the initial field velocity $\Pi_{\rm in}$. As we argue in section~\ref{graceful exit}, 
if $|\phi_{\rm in} - \phi_0 |>| \phi_{\rm in}-\phi_{\rm e}|$ the field performs a graceful exit, with $\phi_{\rm e}$ being eventually reached at all spatial points.
 If, on the other hand, $|\phi_{\rm in} - \phi_0 |<| \phi_{\rm in}-\phi_{\rm e}|$,
 the quantum phase space becomes larger than the classical one such 
 that USR proceeds in two distinct phases, discussed in detail in section~\ref{section:No-graceful-exit}.
 When the quantum particle reaches the end point $\phi_0$
of the classical trajectory, it will start diffusing along the set of classical trajectories 
marked by $\langle\Pi\rangle=0$ and arbitrary $\phi$, implying that the point $\phi_0$ acts as a 
{\it bifurcation point of the quantum phase space}, at which the quantum trajectory splits into 
two branches, see figure~3.1. Consequently, 
a {\it conveyor belt} picture of the quantum particle phase space 
emerges and leads to a phase where some random walkers exit but most are trapped in an eternal de Sitter epoch as the field freely diffuses towards $\phi \rightarrow +\infty$. 
While in the former case a well defined probability distribution for the curvature perturbation emerges, in the latter, although normalizable, the distribution has no finite moments indicating an infinite curvature perturbation. This behavior of course depends on there not being a barrier in reaching 
$\phi \rightarrow +\infty$, a situation not valid in more complete inflationary models.      

This is a preliminary study in many respects. In a realistic inflationary model, the flat potential portion will be finite and even when $|\phi_{\rm in}-\phi_{\rm e}|$ is large and the conveyor belt is operational, the field's diffusion towards large values will be halted, although it may still lead to a greatly enhanced curvature perturbation. Furthermore, in passing to equation~(\ref{PHJ: linearized}) 
we linearized in the field perturbation $\phi-\phi_0$, which is equivalent to assuming a small initial field velocity $\Pi_{\rm in}$. This was done for simplicity and to obtain the semi-analytic results presented here but is not necessary. This paper was largely expository of the methods developed and we will return with a more general treatment and more realistic USR models in a forthcoming publication~\cite{Prokopec:2019ii}. Finally, we dropped the tensor modes which are non-dynamical classically. This statement will no loger hold when their quantum fluctuations are taken into account. We reserve a more sophisticated non-liner treatment of the IR stochastic tensors for the future.

\begin{center}
	{\bf Acknowledgements:}
\end{center} 
GR acknowledges partial supported by the STFC grant ST/P000371/1 – Particles, Fields and Spacetime. 
TP acknowledges the D-ITP consortium, a program of the NWO that is funded by the Dutch Ministry of Education, Culture and Science (OCW). We would like ot thank V. Vennin for very useful discussions and the anonymous referee for insightful comments who allowed us to clarify approximations behind our computations.

\section*{Appendix A: Changing the time-slicing on long wavelengths}

In this appendix, following \cite{Salopek:1990jq}, we recall that under changes of the time hypersurfaces $t\rightarrow T(t,\vc{x})$ the long wavelength equations (\ref{local hubble}), (\ref{H-evolution}), (\ref{Pi-defn}) and  (\ref{Pi-dot}) remain invariant and the long wavelength spatial metric (\ref{3-metric}) retains its form. These statements are valid up to terms which are second order in spatial gradients and are therefore dropped within the long wavelength approximation. 

Starting with coordinates $(t,x^i)$, consider a change in the choice of constant time hyper-surfaces (the spacetime time-slicing) defined by a new time coordinate $T(t,x^i)$. To keep $N_i=0$ in the new coordinate system, new spatial coordinates $X^i$ must also be chosen which are orthogonal to the $T={\rm const}$ surfaces. We now examine how such the transformation between the old and new coordinates can be obtained.

Given the new time surface $T(t,x^i)$, a set of spatial coordinates $X^i$ is chosen on a $T=T_0$ hypersurface and then orthogonally projected to thread all other $T={\rm const}$ hypersurfaces and labelling spatial coordinates in them too. Along constant $X^i$ curves the old coordinates $x^\mu$ will change as $dx^\mu = T^{,\mu}ds $ where $s$ is an arbitrary parameter. Along such lines, $T$ will change as 
\begin{equation}
dT=T_{,\mu}dx^\mu = T_{,\mu}T^{,\mu}ds
\end{equation}                      
which implies
\begin{equation}\label{xmu-T-partial}
\left(\frac{\partial x^\mu}{\partial T}\right)_{X^j}=\frac{T^{,\mu}}{T_{,\alpha}T^{,\alpha}}
\end{equation}
which defines 4 of the 16 components of the transformation matrix between the old and new coordinates. To determine the 12 remaining components consider the transformation matrix 
\begin{equation}
B^\mu_k=\left( \frac{\partial x^\mu}{\partial X^k} \right)_T\,
\end{equation}  
which should be chosen such that 
\begin{equation}\label{orthogonality}
T_{,\mu} B^\mu_k=0\,
\end{equation}
in order to keep $g_{TX^i}=0$. If condition (\ref{orthogonality}) is satisfied on the $T=T_0$ hypersurface it will always be satisfied. This can be seen by taking the $T$ derivative of $B^\mu_k$ to find 
\begin{eqnarray}\label{T-derivative}
\hspace{-0.7cm}\left(\frac{\partial B^\mu_k }{\partial T}\right)_{\!\!X^j} \!\!&=&\! \left(\frac{\partial}{\partial X^k}\left(\frac{\partial x^\mu}{\partial T }\right)_{\!\!X^j}\right)_{\!T} \nonumber\\
&=& \!\left(\frac{\partial}{\partial X^k}\left(\frac{T^{,\mu}}{T_{,\alpha}T^{,\alpha}}\right)\!\right)_{\!T} \!\!=\! B^\nu_k \left(\frac{T^{,\mu}}{T_{,\alpha}T^{,\alpha}}\right)_{\!\!,\nu}
\end{eqnarray} 
In turn, this relation can be used to show that 
\begin{equation}
\partial_T \left(T_{,\mu} B^\mu_k\right)=0
\end{equation}
and hence that $N_i$ is kept zero on all $T$ time-slices in the $\left(T,X^j\right)$ coordinates. 

From (\ref{orthogonality}) we have
\begin{equation}\label{B0k}
B^0_k=-\frac{B^i_k T_{,i}}{T_{,0}}
\end{equation} 
which, when substituted in (\ref{T-derivative}) gives
\begin{equation}
\left(\frac{\partial B^l_k}{\partial T}\right)_{X^j} = \left[\left(\frac{T^{,l}}{T^{,a}T_{,a}}\right)_{,m}-\frac{T_{,m}}{T_{,0}}\left(\frac{T^{,l}}{T^{,a}T_{,a}}\right)_{,0}\right]B^m_k
\end{equation}
The r.h.s.~is second order in spatial gradients and is dropped within our approximation scheme, implying that to this order in the gradient expansion $B^l_k$ is independent of $T$
\begin{equation}\label{B-X-dependence}
B^l_k\equiv B^l_k(\vc{X}). 
\end{equation}
On the other hand, by integrating $x^j$ along a line of constant $X^j$ using (\ref{xmu-T-partial}), we obtain
\begin{equation}
x^j=f^j(\vc{X})+\int \frac{T^{,j}}{T_{,0}T^{,0}} dT 
\end{equation}      
This is consistent with (\ref{B-X-dependence}); a derivative of the second term w.r.t.~$X^j$ involves two spatial gradients $\partial/\partial x^i$ as can be seen by using (\ref{B0k}). Furthermore, any function evaluated at $x^i$ will read
\begin{eqnarray}
g(x^i)&=&g\left(f^i(\vc{X})+\int \frac{T^{,j}}{T_{,0}T^{,0}} dT \right)\nonumber\\ &\simeq& g(f^i(\vc{X})) +g_{,i}\int \frac{T^{,j}}{T_{,0}T^{,0}} dT\,.
\end{eqnarray}     
Hence, within our approximations  
\begin{equation}
g(x^i)= g(f^i(\vc{X})) =\tilde{g}(X^i)
\end{equation} 
and the 3-metric (\ref{3-metric}) in the $(T,\vc{x})$ coordinates reads 
\begin{eqnarray}
\gamma_{l'k'}&=&e^{2\alpha(t,\vc{x})}B^l_{l'}(\vc{X})B^k_{k'}(\vc{X})h_{lk}(\vc{x})\nonumber \\
&=& e^{2\tilde{\alpha}(t(T),\vc{X})}B^l_{l'}(\vc{X})B^k_{k'}(\vc{X})h_{lk}(\vc{X}) 
\end{eqnarray} 
i.e it again has the form of a locally defined conformal factor times a time independent 3-metric which is only a function of the 3 new spatial coordinates $X^i$. The simplest choice for the new spatial coordinates is of course $f^j(\vc{X}) = X^j$.

Regarding the dynamical equations in the new time $T$, we note that for any time dependent quantity $Q$
\begin{equation}
\left(\frac{\partial Q}{\partial T}\right)_{X^j} = \frac{1}{T_{,0}}\left(\frac{\partial Q}{\partial t}\right)_{x^j} + \frac{T^{,k}}{T_{,0}T^{,0}}\left(\frac{\partial Q}{\partial x^k}\right)_{t}\,.
\end{equation}  
Dropping the second term on the r.h.s.~ as second order in spatial gradients and noting that the two lapse functions, $N_t$ and $N_T$, associated with the time coordinates $t$ and $T$ respectively are related by $N_t=N_T \, \partial T/\partial t$, we have
\begin{equation}
\frac{1}{N_T}\left(\frac{\partial Q}{\partial T}\right)_{X^j} = \frac{1}{N_t}\left(\frac{\partial Q}{\partial t}\right)_{x^j} \,.
\end{equation}  
up to second order in spatial gradients. This can be used to show the invariance of the long wavelength dynamical equations under changes of the time slicing.

\section*{APPENDIX B: Steepest descent for $P_{V_0}$}
\label{appendix B}

 The integral in~(\ref{probability on flat branch:2}) is dominated by the dependence on $u$ in
 the exponent, which diverges in both limits of integration, and hence is dominated by 
 some intermediate $u$, at which the function in the exponent minimizes.  
 To study the integral in more detail, we write it in the form, 
 \begin{eqnarray}
 & & P_{V_0} = \frac{3\chi_{\rm in}}{H_0}\left[\mathcal{I}(\chi,\alpha)
                         \!-\!\mathcal{I}(\chi\!-\!2\chi_e,\alpha)\right]
\,,\\
& & \mathcal{I}(\chi,\alpha) = \int_0^\alpha e^{-S(\chi,u)} du
\label{C: integral}
\end{eqnarray} 
where
\begin{eqnarray}
S(\chi,u) &=& \frac{\chi^2}{6(\alpha\!-\!u)}+\frac12\ln\left(\alpha\!-\!u\right)
 -6u \nonumber \\
 &&  \hspace{-0.5cm}-\frac32\ln(2)+n(6u)\chi_{\rm in}^2-\frac32\ln[n(6u)]
 \,,
\label{C: exponential function}
\end{eqnarray}
and where $n(x) = 1/(e^x\!-\!1)$ is the Bose-Einstein function of its argument. 
The integral~(\ref{C: integral}) is then performed by expanding 
$S(\chi,u) $ in~(\ref{C: exponential function}) around the local minimum $u_0$ 
(at which $S^\prime_0\equiv[\partial_u S(\chi,u)]_{u=u_0}=0$)
as, 
\begin{equation}
S(u)\approx S(u_0) + \frac12 S^{\prime\prime}_0(u\!-\!u_0)^2 
      +\mathcal{O}\big((u\!-\!u_0)^3 \big)
\,,
\label{C: expanding exponential function}
\end{equation}
where  $S^{\prime\prime}_0=[\partial_u^2 S(\chi,u)]_{u=u_0}$.
Upon dropping the higher orders $\mathcal{O}\big((u\!-\!u_0)^3 \big)$, the integral~(\ref{C: integral})
becomes simple to evaluate, 
\begin{equation}
\mathcal{I}(\chi,\alpha)=\sqrt{\frac{\pi}{2S^{\prime\prime}_0}} 
  \left[{\rm Erf}\left(\sqrt{\frac{S^{\prime\prime}_0}{2}}(\alpha\!-\!u_0)\right)
        +{\rm Erf}\left(\sqrt{\frac{S^{\prime\prime}_0}{2}}u_0\right)\right]
\end{equation}
where the result is meaningful if $S^{\prime\prime}_0>0$ . To complete the evaluation, we need
$u_0$ and $S^{\prime\prime}_0$, and hence we need the first and second derivative 
of~(\ref{C: exponential function}),
\begin{eqnarray}
 S_0^\prime &=&\frac{\chi^2}{6(\alpha\!-\!u)^2}-\frac{1}{2(\alpha\!-\!u)}-6n(n\!+\!1)\chi_{\rm in}^2+3\!+\!9n\nonumber\\
 &=&0
\label{C: derivative of S 1}\\
  S_0^{\prime\prime} &=&\frac{\chi^2}{3(\alpha\!-\!u)^3}-\frac{1}{2(\alpha\!-\!u)^2}\nonumber\\
&&+36n(n\!+\!1)(n\!+\!2)\chi_{\rm in}^2\!-\!54n(n\!+\!1)
\,,
\label{C: derivative of S 2}
\end{eqnarray}
where we made use of $\partial_u n(6u) = -6n(n\!+\!1)$,  $\partial_u^2 n(6u) = 36n(n\!+\!1)(n\!+\!2)$.
$u_0$ is found by setting $S_0^\prime=0$ in~(\ref{C: derivative of S 1}). 
If $\alpha\gg u$, the problem of solving~(\ref{C: derivative of S 1})
 reduces to finding the positive root of a quadratic equation
in $n=n(6u)$, which is easily solved for $n_0\equiv n(6u_0)$, and hence also for $u_0$.

\end{document}